\documentclass[lettersize,journal]{IEEEtran}
\usepackage{amsmath,amsfonts}
\usepackage{algorithmic}
\usepackage{algorithm}
\usepackage{array}
\usepackage[caption=false,font=footnotesize,labelfont=rm,textfont=rm]{subfig}
\usepackage{booktabs}
\usepackage{makecell}
\usepackage{multirow}
\usepackage{textcomp}
\usepackage{stfloats}
\usepackage{xcolor}
\usepackage{url}
\usepackage{verbatim}
\usepackage{graphicx}
\newtheorem{assumption}{Assumption}
\newtheorem{definition}{Definition} 
\newtheorem{example}{Example}
\newtheorem{theorem}{Theorem}
\newtheorem{lemma}{Lemma}

\usepackage{cite}
\usepackage{tikz}
\usepackage{cancel}
\usepackage[normalem]{ulem}
\usetikzlibrary{positioning}
\usetikzlibrary{arrows.meta, positioning, shapes, fit}
\newcommand{\diffmark}{\textsuperscript{\textcolor{black}{$\ast$}}}

\begin{document}
	
	
	\title{ORBGRAND Is Exactly Capacity-achieving via Rank Companding}
	
	\author{\IEEEauthorblockN{Zhuang Li and Wenyi Zhang},~\IEEEmembership{Senior Member,~IEEE}
		\thanks{The authors are with Department of Electronic Engineering and Information Science, University of Science and Technology of China, Hefei, China (wenyizha@ustc.edu.cn). This work was supported in part by the National Natural Science Foundation of China under Grant 62231022.}
	}
	
	
	
	\maketitle
	
	\begin{abstract}
		Within the family of guessing-based decoding algorithms, ordered reliability bits GRAND (ORBGRAND) has attracted considerable attention due to its efficient use of soft information and suitability for hardware implementation. It has also been shown that ORBGRAND achieves a rate very close to the capacity of an additive white Gaussian noise channel under antipodal signaling. In this work, it is further established that, for general binary-input memoryless channels under symmetric input distribution, via suitably companding the ranks in ORBGRAND according to the inverse cumulative distribution function (CDF) of channel reliability, the resulting CDF-ORBGRAND algorithm exactly achieves the mutual information, i.e., the symmetric capacity. This result is then applied to bit-interleaved coded modulation (BICM) systems to handle high-order input constellations. Via considering the effects of mismatched decoding due to both BICM and ORBGRAND, it is shown that CDF-ORBGRAND is capable of achieving the BICM capacity, which was initially derived in the literature by treating BICM as a set of independent parallel channels.
	\end{abstract}
	
	
	\section{Introduction}\label{sec:intro}
	
	Guessing-based decoding has recently attracted attention as a broad paradigm for channel decoding in emerging communication scenarios. Within this paradigm, guessing random additive noise decoding (GRAND) \cite{duffy2019capacity,riaz2023sub,abbas2023guessing} and its variants form an important line of development, where decoding is performed by progressively guessing the error pattern (EP) experienced by the transmitted codeword. This approach centering around EPs makes the GRAND particularly well suited to short-blocklength or high-rate regimes \cite{wang25arxiv}.    
	
	GRAND was originally proposed for hard-decision channels. This is also reflected by its name: only when the noise is discrete it makes sense to ``guess'' it. For more general channels, such as additive white Gaussian noise (AWGN) channels, incorporating channel soft information can improve decoding performance \cite[Ch. 10]{Lin2004ErrorCC}, and we need to guess the EP rather than the noise itself. Symbol reliability GRAND (SRGRAND) \cite{duffy2021guessing} uses a single-bit reliability indicator to mark whether a channel output symbol is reliable. Soft GRAND (SGRAND)\cite{solomon2020soft} directly exploits the exact log-likelihood ratio (LLR) magnitudes of the channel output to determine the sequence of EPs for guessing, thereby fully leveraging the soft information and achieving maximum-likelihood (ML) decoding. However, the requirement of exact LLR magnitudes in SGRAND necessitates on-the-fly generation of EPs, posing challenges for hardware implementation. Another variant of GRAND, ordered reliability bits GRAND (ORBGRAND) \cite{duffy2022ordered} generates EPs based on the ranking of LLR magnitudes rather than their exact values. This approach enables efficient generation of EPs without depending upon exact LLR magnitudes, and streamlines hardware implementation \cite{condo2021high,abbas2022high,condo2022fixed}. It has also been shown in \cite{liu2022orbgrand} through an information-theoretic analysis that for AWGN channels under antipodal signaling, the rate achieved by ORBGRAND is very close to the channel mutual information, i.e., the capacity in that setup. There also exist several works that treat GRAND for more general channels such as fading channels \cite{sarieddeen2022grand,abbas2022grand,abbas2023hardware,chatzigeorgiou2022symbol,allahkaram2022urllc,allahkaram2025constrained} and impulsive noise channels \cite{feng2024laplacian,wiame2024error}.
	
	In this paper, we analyze a modification of ORBGRAND, which essentially adopts a companding technique to transform the rank of each LLR magnitude according to the inverse cumulative distribution function (CDF) of LLR magnitude. The resulting algorithm is thus termed CDF-ORBGRAND. The conceptual motivation for CDF-ORBGRAND can be traced back to the original ORBGRAND\cite{duffy2022ordered}, where decoding is based on the ranking of channel reliabilities and is naturally connected to an order-statistics view of the reliability sequence. In particular, the effectiveness of ORBGRAND suggests that much of the benefit of channel soft information can be captured through the ranks of channel reliabilities, without requiring their exact values. This line of understanding was later made more explicit in \cite{liu2022orbgrand}, where cdf-GRAND was introduced as an asymptotically equivalent proxy for understanding ORBGRAND. From that viewpoint, for large blocklengths, the normalized rank of a reliability magnitude approximates its CDF value, so that applying the inverse CDF naturally maps the rank back to the reliability scale. The value of the shape of the CDF of LLR magnitude has also been recognized earlier in \cite{duffy2022ordered},  wherein a piecewise linear approximation technique has been utilized under the term of ``full ORBGRAND''; see also \cite{wan2024approaching,wan2025fine}. In contrast to cdf-GRAND, which still relies on the exact LLR magnitudes, CDF-ORBGRAND applies the inverse CDF directly to normalized ranks. Since the inverse-CDF can be evaluated offline, CDF-ORBGRAND retains the rank-based structure of ORBGRAND and its corresponding hardware-friendly implementation, while asymptotically aligning the resulting decoding metric with the channel law. The main contribution of this paper lies in the rigorous achievable rate characterization of CDF-ORBGRAND, in proving that this rank-based decoding rule, although mismatched, is exactly capacity-achieving for general binary-input memoryless channels, and in extending this result to bit-interleaved coded modulation (BICM) systems, where CDF-ORBGRAND is shown to achieve the BICM capacity.
	
	Specifically, we analyze the achievable rate of CDF-ORBGRAND for general binary-input memoryless channels under symmetric input distribution, and prove via a random coding argument akin to that in \cite{liu2022orbgrand} that, CDF-ORBGRAND exactly achieves the mutual information, i.e., the symmetric capacity, without any rate loss. While numerical results suggest that ORBGRAND already operates very close to the symmetric capacity for several representative channels, such observations do not resolve the more fundamental question of whether a rank-based decoding rule can be exactly capacity-achieving. Our result shows that CDF-ORBGRAND achieves the symmetric capacity exactly. Thus, beyond the practical relevance of GRAND and its variants in short-blocklength and high-rate regimes, at the level of achievable rate, restricting the decoder to exploit rank-based reliability information only does not necessarily incur any rate loss, provided that an appropriate CDF-based companding is applied. In this sense, the present work establishes the achievable rate limit of a class of rank-based GRAND decoders, further sheds light on the behavior of ORBGRAND, and may also provide a useful theoretical basis for future finite-blocklength studies of ORBGRAND and related rank-based variants.
	
	On the basis of CDF-ORBGRAND for binary-input channels, we proceed to the investigation of BICM systems \cite{caire1998bit,i2008bit}. This investigation is motivated by the prevalence of high-order modulation in modern communication systems, wherein BICM has been a widely used technique. By jointly considering the mismatched decoding effects due to both BICM and ORBGRAND, we derive achievable rates of both CDF-ORBGRAND and ORBGRAND for BICM. Remarkably, we show that CDF-ORBGRAND achieves the BICM capacity, which was initially derived in the literature by treating BICM as a set of independent parallel channels \cite{caire1998bit}, and subsequently established by typicality decoding or by mismatched decoding with appropriately chosen metrics \cite{martinez2009bit}. The BICM capacity-achieving property of CDF-ORBGRAND indicates that the rank companding technique is effective even for handling the heterogeneity among bit channels corresponding to different positions of the labeling of constellation symbols. 
	
	The remaining part of this paper is organized as follows: Section \ref{sec:preliminaries} introduces the system model, along with GRAND and its variants for general binary-input memoryless channels. Section \ref{sec:rate-cdf-orb} proves that CDF-ORBGRAND exactly achieves the symmetric capacity. Section \ref{sec:bicm} considers applying ORBGRAND and CDF-ORBGRAND to a BICM system model, and derives their respective achievable rates, revealing that CDF-ORBGRAND achieves the BICM capacity. Section \ref{Conclusion} concludes this paper. Technical proofs are provided in Appendices.
	
	\section{Preliminaries}\label{sec:preliminaries}
	Throughout the paper, symbols in sans-serif font, such as $\mathsf{X}$, denote random variables, and the corresponding ordinary symbols, such as $x$, denote their realizations. Underlined symbols denote vectors; for example, $\underline{\mathsf{Y}}$ and $\underline{y}$ denote a random vector and its realization, respectively, whereas $y$ denotes a scalar symbol and $y_i$ denotes the $i$-th element of $\underline{y}$. A summary of the main symbols used below is provided in Table~\ref{tab:notation}.
	\begin{table}[htbp]
		\centering
		\caption{Main symbols used in the system model.}
		\label{tab:notation}
		\begin{tabular}{c|l}
			\hline
			Symbol & Meaning \\
			\hline
			$\mathsf{W}$   & transmitted message \\
			$\mathsf{X}$   & channel input \\
			$\mathsf{Y}$   & channel output \\
			$\mathsf{T}$   & log-likelihood ratio (LLR) \\
			$\mathsf{R}_i$ & rank of $|\mathsf{T}_i|$ among the channel reliability magnitudes \\
			\hline
		\end{tabular}
	\end{table}
	
	\subsection{System Model}\label{subsec:system model}
	
	Our basic channel model is a general binary-input memoryless channel, which, without loss of generality, is of input alphabet $\{+1, -1\}$, with the output probability density function (pdf) denoted by $q^+(y)$ when input $x = +1$, and by $q^-(y)$ when input $x = -1$. Here $q^+(y)$ and $q^-(y)$ are general, without assuming any specific structure like symmetry.
	
	A codebook has length $N$ and rate $R$ nats per channel use, corresponding to $\lceil e^{NR}\rceil$ possible messages. When transmitting message $w$, the corresponding codeword is denoted by $\underline{x}(w) = [x_1(w),\cdots,x_N(w)]$. In this paper we conduct a random coding analysis, and hence assume that the elements of $\underline{x}(w)$ are independent and identically distributed (i.i.d.), uniformly drawn from $\{+1, -1\}$. A message $\mathsf{W}$ is uniformly selected from $\{1, \ldots, \lceil e^{NR}\rceil\}$ for transmission. Given the channel output vector $\underline{\mathsf{Y}} = [\mathsf{Y}_1,\cdots,\mathsf{Y}_N]$, we define the LLRs as $$\mathsf{T}_i=\ln\frac{q^+(\mathsf{Y}_i)}{q^-(\mathsf{Y}_i)}, \quad i=1,\cdots,N,$$
	and call their magnitudes, $[\lvert \mathsf{T}_1\rvert,\cdots,\lvert \mathsf{T}_N\rvert]$, the channel reliability vector. The CDF of $\lvert \mathsf{T}\rvert$ is denoted by $\Psi(t)$, for $t\geq 0$. We also introduce the sets $\mathcal{R}_1=\left\{y\vert q^+(y)<q^-(y)\right\}$ and $\mathcal{R}_2=\left\{y\vert q^+(y)>q^-(y)\right\}$.
	
	For $i = 1,\cdots,N$, we denote $\mathsf{R}_i$ as the rank of $\lvert\mathsf{T}_i\rvert$ among the channel reliability vector, from $1$ (the smallest) to $N$ (the largest).
	
	For our analysis, we impose a few additional assumptions on the behavior of $q^+(y)$ and $q^-(y)$, as well as on the induced reliability CDF $\Psi(\cdot)$ and its inverse, as follows.
	\begin{assumption}\label{assumption:light-tail}
		Both $q^+(y)$ and $q^-(y)$ are light-tailed, in the sense that there exists a sufficiently large constant $M_1>0$, a constant $a>0$ and a polynomial $S_1(\cdot)$ satisfying
		\begin{align}
			q^{\pm}(y)<S_1(\lvert y \rvert)e^{-a\lvert y\rvert}, \quad \forall \lvert y \rvert > M_1.
		\end{align}
	\end{assumption}
	
	\begin{assumption}\label{assumption:poly-bound}
		The growth of $\lvert t\rvert=\bigl\lvert\ln\frac{q^+(y)}{q^-(y)}\bigr\rvert$ is polynomially bounded; that is, there exists a sufficiently large constant $M_2>0$ and a polynomial $S_2(\cdot)$ satisfying $\lvert t\rvert<S_2(\lvert y \rvert), \forall \lvert y\rvert>M_2$.
	\end{assumption}
	
	\begin{assumption}\label{assumption:finite-derivative}
		The CDF $\Psi(\cdot)$ of $|\mathsf{T}|$ is continuous. Throughout this paper, 
		$\Psi^{-1}(\cdot)$ denotes the inverse of $\Psi(\cdot)$, defined by
		\begin{align}
			\Psi^{-1}(u)=\inf\{t\geq0:\Psi(t)>u\},\quad 0\leq u<1.
		\end{align}
		Moreover, for every finite reliability value $v$ in the support of $|\mathsf{T}|$ with $\Psi(v)<1$, 
		\begin{align}
			\Psi^{-1}(\Psi(v))=v,
		\end{align}
		and $\Psi^{-1}(\cdot)$ is continuous at $\Psi(v)$. \footnote{In statistics, $\Psi^{-1}(\cdot)$ is also called the quantile function of $\lvert\mathsf{T}\rvert$.}
	\end{assumption}
	
	\subsection{GRAND and Its Variants}\label{subsec:grand}
	
	Upon receiving a channel output vector $\underline{y}$, we compute the channel reliability vector $\left[\lvert t_1\rvert,  \cdots,\lvert t_N\rvert\right]$ and the corresponding hard-decision vector $\underline{x}_{\text{hard}}=[\text{sgn}(t_1), \cdots, \text{sgn}(t_N)]$, where $t_i=\ln\frac{q^+(y_i)}{q^-(y_i)}$ is the LLR and $\text{sgn}(t) = 1$ if $t \geq 0$ and $-1$ otherwise.
	
	In general, we can collectively denote the EPs used by GRAND as a $2^N\times N$ $\pm 1$-valued matrix $P$: in the $q$-th query, the $q$-th row of $P$ is used as the EP for testing, such that if $P_{q, i} = -1$ the sign of the $i$-th entry of $\underline{x}_{\text{hard}}$ is flipped, and otherwise it remains unchanged. If $\underline{x}_{\text{hard}}$ after flipping is a codeword, it is declared as the decoder output. If all rows of $P$ are exhausted without finding a codeword, a decoding failure is declared.
	
	For ORBGRAND, we arrange the rows of $P$ so that the sum reliability of the $q$-th row, $\sum_{i: P_{q, i} = -1} r_i$, is non-decreasing with $q$, where $r_i$ is the rank of $\lvert t_i\rvert$ introduced in the previous subsection. Efficient algorithms for generating such $P$ are available (see, e.g., \cite{duffy2022ordered, condo2021high}). Since $2^N$ is typically an exceedingly large quantity, in practice $P$ is usually truncated to its first $Q$ rows, with $Q$ being the maximum number of queries allowed; for our information-theoretic analysis in this paper, however, we do not consider the effect of truncation and hence all EPs of $P$ will be queried in the worst case.
	
	As shown in \cite[Sec. II]{liu2022orbgrand}, ORBGRAND and many other variants of GRAND can be represented by the following unified decoding rule, as long as no truncation of $P$ is applied, i.e., $Q = 2^N$:
	\begin{align}
		\label{decoding rule}
		\hat{w} = \mathop{\arg\min}\limits_{w = 1,\cdots, \lceil e^{NR}\rceil}\frac{1}{N}\sum_{i=1}^{N} \gamma_i(\underline{y})\cdot \mathbf{1}(\text{sgn}(t_i )\cdot x_i(w)< 0 ).
	\end{align}
	Different choices of $\left\{\gamma_i\right\}_{i=1,\cdots,N}$ correspond to different choices of $P$ and thus different variants of GRAND:
	\begin{itemize}
		\item When $\gamma_i(\underline{y})=1$, (\ref{decoding rule}) is the original GRAND \cite{duffy2019capacity};
		\item When $\gamma_i(\underline{y})=\left\lvert t_i\right\rvert$, (\ref{decoding rule}) is SGRAND \cite{solomon2020soft}, which is equivalent to ML decoding;
		\item When $\gamma_i(\underline{y})=\frac{r_i}{N}$, (\ref{decoding rule}) is ORBGRAND \cite{duffy2022ordered};
		\item When $\gamma_i(\underline{y})=\Psi(\left\lvert t_i\right\rvert)$, (\ref{decoding rule}) is cdf-GRAND \cite{liu2022orbgrand};
		\item When $\gamma_i(\underline{y})=\Psi^{-1}\left(\frac{r_i}{N+1}\right)$, where $\Psi^{-1}(\cdot)$ denotes the inverse function of the CDF of $\lvert\mathsf{T}\rvert$, $\Psi(\cdot)$, (\ref{decoding rule}) is called CDF-ORBGRAND. Here, the normalized rank $\frac{r_i}{N+1}$ is transformed through the inverse CDF $\Psi^{-1}(\cdot)$, thereby transforming it from the rank domain to the reliability domain. In this paper, the term ``companding'' refers to this nonlinear inverse-CDF transformation.
	\end{itemize}
	
	The form of (\ref{decoding rule}) facilitates information-theoretic analysis, as will be conducted in the following sections. A decoding error event occurs if $\hat{\mathsf{W}} \neq \mathsf{W}$, where $\hat{\mathsf{W}}$ is the decoder output, and we define the achievable rate of GRAND as follows.
	\begin{definition}
		For the system model in Section \ref{subsec:system model}, a rate $R$ is achievable under GRAND with given $\{\gamma_i\}_{i = 1, \ldots, N}$ if there exists a sequence of codebooks of length $N = 1, 2, \ldots$ , such that the average error probability of decoding for the decoding rule (\ref{decoding rule}) asymptotically vanishes as $N$ grows without bound.
	\end{definition}
	
	If we do not restrict the decoding rule to be (\ref{decoding rule}) and apply ML decoding, then for our system model in Section \ref{subsec:system model}, according to Shannon's channel coding theorem, the supremum of all achievable rates is the mutual information $I(\mathsf{X}; \mathsf{Y})$ with $\mathsf{X}$ uniformly distributed over $\{+1, -1\}$. This mutual information thus serves as the ultimate performance limit for the considered system model, and is termed the symmetric capacity in the sequel.
	
	\section{Capacity-achieving Property of CDF-ORBGRAND}\label{sec:rate-cdf-orb}
	
	To motivate CDF-ORBGRAND, it is helpful to inspect the decoding rule (\ref{decoding rule}). Based on the convergence of empirical distributions, as $N$ gets large, with high probability, $\mathsf{R}_i/(N + 1)$ gets close to $\Psi(\lvert\mathsf{T}_i\rvert)$. Thus, by conducting the inverse operation, i.e., $\Psi^{-1}(\cdot)$, over $\mathsf{R}_i/(N + 1)$, we essentially return to $\lvert\mathsf{T}_i\rvert$, and this is exactly what CDF-ORBGRAND does. Due to this intuitive argument, CDF-ORBGRAND should behave similarly to SGRAND asymptotically, and this assertion will be formalized and established in this section. Since $\left\{\gamma_i\right\}_{i=1,\cdots,N}$ simply compands the ranks $\left\{\mathsf{R}_i\right\}_{i=1,\cdots,N}$, the corresponding EP queries can also be generated in a similar fashion to those of ORBGRAND. 
	
	We use the following example to illustrate how the guessing order induced by CDF-ORBGRAND compares with those induced by SGRAND and ORBGRAND at a short blocklength where a partial EP ordering can be explicitly displayed.
	
	\begin{example}\label{exa1}
		Consider a BPSK-modulated AWGN channel
		\[
		\mathsf{Y}=\sqrt{P}\mathsf{X}+\mathsf{Z},
		\]
		where $\mathsf{Z}\sim\mathcal{N}(0,1)$, so that the channel LLR is $\mathsf{T}=2\sqrt{P}\mathsf{Y}$. Let  $N=5$ and the signal-to-noise ratio (SNR) be $6$ dB. Suppose that the reliability vector is  $[|t_1|,|t_2|,|t_3|,|t_4|, |t_5| ]=[5.17, 6.08, 7.93, 9.56, 12.01]$, whose corresponding ranks are  $\underline{r}=[1,2,3,4,5]$. The corresponding companded values used by CDF-ORBGRAND are $[4.12, 6.25, 7.96, 9.68, 11.82]$.
		To make the guessing functions explicit, let  $\underline{e}=[e_1,e_2,e_3,e_4,e_5]\in\{+1,-1\}^5$ denote an EP. Under this representation, a negative entry $e_i=-1$ indicates that the $i$-th element of the hard-decision vector is flipped by the EP, while $e_i=+1$ indicates that it remains unchanged. This follows naturally from the general framework in~(\ref{decoding rule}). Then SGRAND, ORBGRAND, and CDF-ORBGRAND rank EPs according to
		\begin{align*}
			&\sum_{i=1}^N |t_i|\cdot\mathbf{1}(e_i<0),
			\quad
			\sum_{i=1}^N r_i\cdot\mathbf{1}(e_i<0) ,\\
			&\quad\text{and}\quad
			\sum_{i=1}^N \Psi^{-1}\left(\frac{r_i}{N+1}\right)\cdot\mathbf{1}(e_i<0)
			,
		\end{align*}
		respectively. The first twenty EPs in the resulting guessing orders, together with the corresponding metric values, are listed in Table~\ref{tab:new_order_example}. For ORBGRAND, when multiple EPs have the same metric value, their relative order may be chosen arbitrarily; the ordering shown in Table~\ref{tab:new_order_example} follows one such choice, where ties are broken to favor the SGRAND ordering. Compared with ORBGRAND, CDF-ORBGRAND induces an EP ordering much closer to that of SGRAND. In the first twenty EPs listed in Table~\ref{tab:new_order_example}, ORBGRAND differs from SGRAND in ten positions, whereas CDF-ORBGRAND differs only by an adjacent swap between Orders 9 and 10. Entries marked with $^\ast$ indicate differences from the SGRAND ordering.
	\end{example}
	
	\begin{table}[htbp]
		\centering
		\scriptsize
		\setlength{\tabcolsep}{4pt}
		\caption{Partial EP ordering and corresponding metric values in Example~\ref{exa1}, where \texttt{+} and \texttt{-} denote $+1$ and $-1$, respectively.}
		\label{tab:new_order_example}
		\begin{tabular}{c|cc|cc|cc}
			\toprule
			\multirow{2}{*}{Order} 
			& \multicolumn{2}{c|}{SGRAND} 
			& \multicolumn{2}{c|}{ORBGRAND} 
			& \multicolumn{2}{c}{CDF-ORBGRAND} \\
			& EP & Metric & EP & Metric & EP & Metric \\
			\midrule
			1  & \texttt{+++++} & $0$     & \texttt{+++++} & $0$ & \texttt{+++++} & $0$ \\
			2  & \texttt{-++++} & $5.17$  & \texttt{-++++} & $1$ & \texttt{-++++} & $4.12$ \\
			3  & \texttt{+-+++} & $6.08$  & \texttt{+-+++} & $2$ & \texttt{+-+++} & $6.25$ \\
			4  & \texttt{++-++} & $7.93$  & \texttt{++-++} & $3$ & \texttt{++-++} & $7.96$ \\
			5  & \texttt{+++-+} & $9.56$  & \texttt{--+++}\diffmark & $3$ & \texttt{+++-+} & $9.68$ \\
			6  & \texttt{--+++} & $11.25$ & \texttt{+++-+}\diffmark & $4$ & \texttt{--+++} & $10.37$ \\
			7  & \texttt{++++-} & $12.01$ & \texttt{-+-++}\diffmark & $4$ & \texttt{++++-} & $11.82$ \\
			8  & \texttt{-+-++} & $13.10$ & \texttt{++++-}\diffmark & $5$ & \texttt{-+-++} & $12.08$ \\
			9  & \texttt{+--++} & $14.01$ & \texttt{+--++} & $5$ & \texttt{-++-+}\diffmark & $13.80$ \\
			10 & \texttt{-++-+} & $14.73$ & \texttt{-++-+} & $5$ & \texttt{+--++}\diffmark & $14.21$ \\
			11 & \texttt{+-+-+} & $15.64$ & \texttt{+-+-+} & $6$ & \texttt{+-+-+} & $15.93$ \\
			12 & \texttt{-+++-} & $17.18$ & \texttt{-+++-} & $6$ & \texttt{-+++-} & $15.94$ \\
			13 & \texttt{++--+} & $17.49$ & \texttt{---++}\diffmark & $6$ & \texttt{++--+} & $17.64$ \\
			14 & \texttt{+-++-} & $18.09$ & \texttt{+-++-} & $7$ & \texttt{+-++-} & $18.07$ \\
			15 & \texttt{---++} & $19.18$ & \texttt{++--+}\diffmark & $7$ & \texttt{---++} & $18.33$ \\
			16 & \texttt{++-+-} & $19.94$ & \texttt{--+-+}\diffmark & $7$ & \texttt{++-+-} & $19.78$ \\
			17 & \texttt{--+-+} & $20.81$ & \texttt{++-+-}\diffmark & $8$ & \texttt{--+-+} & $20.05$ \\
			18 & \texttt{+++--} & $21.57$ & \texttt{--++-}\diffmark & $8$ & \texttt{+++--} & $21.50$ \\
			19 & \texttt{-+--+} & $22.66$ & \texttt{-+--+} & $8$ & \texttt{-+--+} & $21.76$ \\
			20 & \texttt{--++-} & $23.26$ & \texttt{+++--}\diffmark & $9$ & \texttt{--++-} & $22.19$ \\
			\bottomrule
		\end{tabular}
	\end{table}

	The next example complements the previous one by illustrating, for a larger blocklength, why the companded metric used by CDF-ORBGRAND more closely tracks the true reliabilities than ORBGRAND.
	\begin{figure}[htbp]
		\centering
		\includegraphics[width=0.49\textwidth]{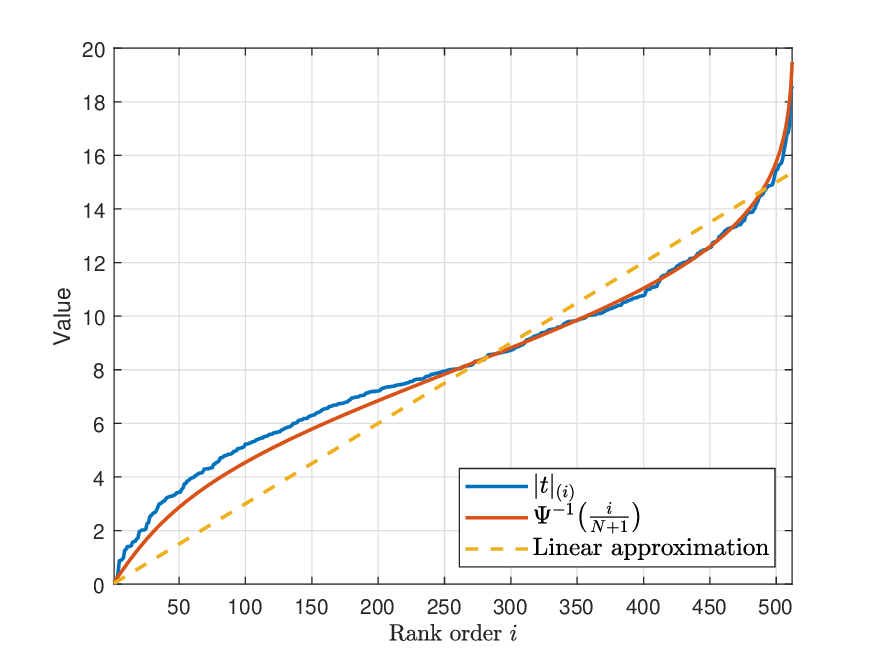}
		\caption{Comparison of the channel reliabilities, the companded values, and the linear approximation for a 512-bit sample over the BPSK-modulated AWGN channel with $\text{SNR}=6$ dB.}
		\label{fig:example2}
	\end{figure}
	\begin{example}
		Fig. \ref{fig:example2} shows a sample 512-bit sequence. After sorting the channel reliabilities in ascending order, we plot the corresponding values of $|t|_{(i)}$, $\Psi^{-1}\left(\frac{i}{N+1}\right)$, and their linear approximation, which correspond to the metrics used by SGRAND, CDF-ORBGRAND, and ORBGRAND, respectively; see, e.g., \cite{duffy2022ordered} for details on the linear approximation underlying ORBGRAND. Here, $|t|_{(i)}$ denotes the $i$-th smallest value among $\{|t_1|,\ldots,|t_N|\}$. It can be seen that CDF-ORBGRAND provides a much more accurate approximation to the true channel reliabilities than ORBGRAND. Consequently, the EP ordering induced by CDF-ORBGRAND is also closer to the optimal ordering, namely, that induced by SGRAND.
	\end{example}

	\subsection{Main Result}\label{subsec:rate-cdf-orb}
	
	Neither ORBGRAND nor CDF-ORBGRAND is ML decoding, and furthermore, their metrics, $\left\{\gamma_i\right\}_{i=1,\cdots,N}$, are $N$ correlated random variables due to ranking. Therefore, we cannot directly invoke existing information-theoretic formulas for evaluating their achievable rates. In this paper, we essentially establish the achievable rate of CDF-ORBGRAND as the generalized mutual information (GMI) of the considered decoding rule, following the general approach of mismatched decoding (see, e.g., \cite{ganti2000mismatched, lapidoth2002fading}). As revealed by the result below, although CDF-ORBGRAND is inherently mismatched, its achievable rate exactly coincides with the symmetric capacity. In other words, from an information-theoretic perspective, no rate loss is incurred when applying CDF-ORBGRAND.


	\begin{theorem}
		\label{thm:gmi}
		For the system model in Section \ref{subsec:system model}, CDF-ORBGRAND achieves the symmetric capacity, i.e.,
		\begin{align}
			I(\mathsf{X};\mathsf{Y}) &=\ln2-\frac{1}{2}\int_{-\infty}^{\infty}\ln\left(1+\frac{q^-(y)}{q^+(y)}\right)q^+(y)\mathrm{d}y\nonumber\\
			&\hspace{3em}-\frac{1}{2}\int_{-\infty}^{\infty}\ln\left(1+\frac{q^+(y)}{q^-(y)}\right)q^-(y)\mathrm{d}y
			\label{channel mutual information}
		\end{align}
		in nats/channel use.
	\end{theorem}
	\begin{IEEEproof}
		See Section \ref{subsec:proof-cdf-rate}.
	\end{IEEEproof}
	
	In our previous work\cite{li2024orbgrand}, we have derived the GMI of ORBGRAND for the system model in Section \ref{subsec:system model} as
	\begin{align}
		I_{\text{ORB}} &= \ln2 - \inf\limits_{\theta < 0}\Bigg\{\int_{0}^{1} \ln(1 + e^{\theta t})\mathrm{d}t\nonumber\\
		&\hspace{4em}-\theta\cdot\frac{1}{2}\int_{\mathcal{R}_1}\Psi\left(\Bigl\lvert\ln\frac{q^+(y)}{q^-(y)}\Bigr\rvert\right) q^+(y)\mathrm{d}y\nonumber\\
		&\hspace{4em}-\theta\cdot\frac{1}{2}\int_{\mathcal{R}_2}\Psi\left(\Bigl\lvert\ln\frac{q^+(y)}{q^-(y)}\Bigr\rvert\right)q^-(y)\mathrm{d}y\Bigg\}
		\label{ORBGRAND rate}
	\end{align}    
	in nats/channel use. According to the proof of Theorem \ref{thm:gmi} in Section \ref{subsec:proof-cdf-rate}, the symmetric capacity (\ref{channel mutual information}) can be equivalently rewritten as
	\begin{align}
		I_{\text{CDF-ORB}} &= \ln2 - \inf\limits_{\theta < 0}\Bigg\{\mathbb{E}\left[\ln\left(1+e^{\theta\cdot\bigl\lvert\ln\frac{q^+(\mathsf{Y})}{q^-(\mathsf{Y})}\bigr\rvert}\right)\right]\nonumber\\
		&\quad-\theta\cdot\frac{1}{2}\int_{\mathcal{R}_1}\Bigl\lvert \ln\frac{q^+(y)}{q^-(y)}\Bigr\rvert q^+(y)\mathrm{d}y\nonumber\\
		&\quad-\theta\cdot\frac{1}{2}\int_{\mathcal{R}_2}\Bigl\lvert \ln\frac{q^+(y)}{q^-(y)}\Bigr\rvert q^-(y)\mathrm{d}y\Bigg\}
		\label{CDF-ORBGRAND rate}
	\end{align} 
	in nats/channel use. Comparing (\ref{ORBGRAND rate}) and (\ref{CDF-ORBGRAND rate}), we observe that their difference lies in that for CDF-ORBGRAND the integrands are $\bigl\lvert \ln\frac{q^+(y)}{q^-(y)}\bigr\rvert$ and for ORBGRAND the corresponding integrands are replaced by $\Psi\Bigl(\bigl\lvert \ln\frac{q^+(y)}{q^-(y)}\bigr\rvert\Bigr)$.\footnote{It can be readily shown that with this replacement the first expectation term in the bracket of (\ref{CDF-ORBGRAND rate}) becomes the integral $\int_{0}^{1} \ln(1 + e^{\theta t})\mathrm{d}t$ of (\ref{ORBGRAND rate}), since $\Psi\left(\left\lvert \mathsf{T} \right\rvert\right)$ is a uniform random variable over the unit interval.}
	
\begin{figure}[htbp]	
		\centering
		\includegraphics[width=0.49\textwidth]{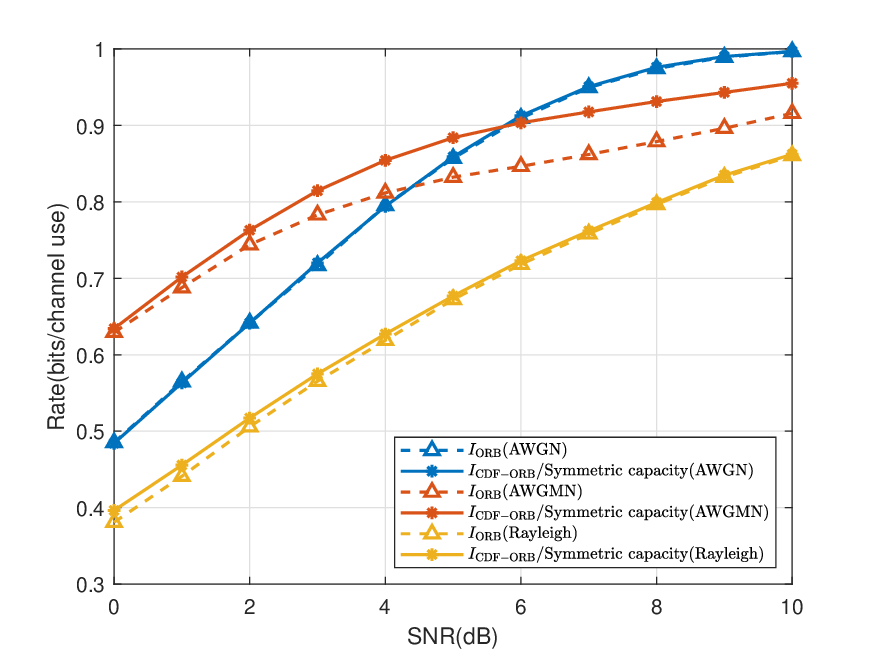}
		\caption{Comparison between $I_{\text{ORB}}$ and $I_{\text{CDF-ORB}}$ under AWGN, AWGMN, and Rayleigh fading channels. Since Rayleigh fading channels are described in complex baseband, the SNR is defined as the effective per-dimension SNR, corresponding to the in-phase component of the received signal.}
		\label{fig:rate}
	\end{figure}

	We conduct some numerical experiments to compare (\ref{ORBGRAND rate}) and (\ref{CDF-ORBGRAND rate}), for antipodal signaling, i.e., BPSK modulation, over AWGN,  additive white Gaussian-mixture noise (AWGMN), and Rayleigh fading channels with receiver-side perfect channel state information (CSI). These channels can be written as
	\begin{equation}
		\mathsf{Y} = \sqrt{P} \mathsf{H}\mathsf{X} + \mathsf{Z}:\nonumber
	\end{equation}
	\begin{itemize}
		\item AWGN channel, $\mathsf{H}=1$ and $\mathsf{Z}\sim\mathcal{N}(0,1)$: $\mathsf{T} = 2\sqrt{P}\mathsf{Y}$.
		\item AWGMN channel, $\mathsf{H}=1$ and 
			$\mathsf{Z}\sim \sum_{\ell=1}^{L}\alpha_\ell \mathcal{N}(0,\sigma_\ell^2)$, 
			where $\alpha_\ell>0$, $\sum_{\ell=1}^{L}\alpha_\ell=1$, and $\sigma_\ell^2>0$. 
			For the two-component mixture used in the numerical results, the corresponding LLR is
			\[
			\mathsf{T}
			=
			\ln
			\frac{
				\sum_{\ell=1}^{2}\alpha_\ell
				\phi(\mathsf{Y}-\sqrt{P};0,\sigma_\ell^2)
			}{
				\sum_{\ell=1}^{2}\alpha_\ell
				\phi(\mathsf{Y}+\sqrt{P};0,\sigma_\ell^2)
			},
			\]
			where $\phi(\cdot;0,\sigma^2)$ denotes the pdf of $\mathcal{N}(0,\sigma^2)$. 
			Specifically, we set $\alpha_1=0.95$, $\alpha_2=0.05$, 
			$\sigma_1^2=10/19$, and $\sigma_2^2=10$, so that $\mathbb{E}[\mathsf{Z}^2]=1$.
		\item Rayleigh fading channel with receiver-side perfect CSI, $\mathsf{H}\sim\mathcal{CN}(0,1)$ and $\mathsf{Z}\sim\mathcal{CN}(0,1)$: $\mathsf{T} = 4\sqrt{P}\Re(\mathsf{H}^*\mathsf{Y})$. We point out that here $q^\pm(\cdot)$ should be evaluated for the augmented channel output $(y, h)$, i.e., including the receiver-side perfect CSI.\footnote{The program code can be found at: \url{https://github.com/ustclizhuang/Simulation}}
	\end{itemize}
	We note that $I_{\text{CDF-ORB}}$ in Theorem \ref{thm:gmi} and $I_{\text{ORB}}$ in (\ref{ORBGRAND rate}) are measured in nats rather than bits. In numerical plots, we convert them into bits for convenience. The curves of $I_{\text{ORB}}$ and $I_{\text{CDF-ORB}}$ for different SNR values are displayed in Fig.~\ref{fig:rate}. Their comparison indicates that in AWGN and Rayleigh fading channels, the rate loss of ORBGRAND is essentially negligible, an observation already made in \cite{liu2022orbgrand,li2024orbgrand},  and that in AWGMN channel, the gap becomes more pronounced in the moderate-to-high SNR regime.
	
	\subsection{Proof of Main Result}\label{subsec:proof-cdf-rate}
	
	Based on the unified decoding rule (\ref{decoding rule}), the decoding metric of CDF-ORBGRAND is given by
	\begin{align}
		\mathsf{D}(w)=\frac{1}{N}\sum_{i=1}^{N} \Psi^{-1}\left(\frac{\mathsf{R}_i}{N+1}\right)\cdot \mathbf{1}\left(\text{sgn}\left(\mathsf{T}_i\right)\cdot \mathsf{X}_i(w)< 0\right),
		\label{decoding metric}
	\end{align}
	for $w=1, \cdots, \lceil e^{NR}\rceil$.
	
	Since the terms in the summation in (\ref{decoding metric}) involve ranks of channel reliabilities, they are correlated, and hence the standard GMI formula (see, e.g., \cite[Eqn. (12)]{ganti2000mismatched}) cannot be straightforwardly applied. Therefore we conduct the analysis directly from the first principle, akin to that in \cite{liu2022orbgrand,li2024orbgrand} and dating back to \cite{lapidoth2002fading,weingarten04it}. We calculate the ensemble average error probability of decoding, which, for the i.i.d. random coding ensemble in our system model, is equivalent to the error probability of decoding when message $w=1$ is transmitted.
	
	Under the condition that the transmitted message is $w=1$, some asymptotic properties of the CDF-ORBGRAND decoding metric (\ref{decoding metric}) are presented in the following lemmas, whose proofs are provided in the appendices.
	\begin{lemma}
		\label{lemma CDF-ORB exp}
		As $N \rightarrow \infty$, the expectation of the decoding metric (\ref{decoding metric}) for the transmitted message $w=1$ is given by
		\begin{align}
			\lim\limits_{N\to\infty} \mathbb{E} \mathsf{D}(1) &= \frac{1}{2}\int_{\mathcal{R}_1}\Bigl\lvert \ln\frac{q^+(y)}{q^-(y)}\Bigr\rvert q^+(y)\mathrm{d}y\nonumber\\
			&\quad +\frac{1}{2}\int_{\mathcal{R}_2}\Bigl\lvert \ln\frac{q^+(y)}{q^-(y)}\Bigr\rvert q^-(y)\mathrm{d}y.  
		\end{align}     	
	\end{lemma}
	\begin{lemma}
		\label{lemma CDF-ORB variance}
		As $N \rightarrow \infty$, the variance of the decoding metric (\ref{decoding metric}) for the transmitted message $w=1$ satisfies
		\begin{align}
			\lim\limits_{N\to\infty} \text{var} \mathsf{D}(1) =  0.          
		\end{align}
	\end{lemma}
	\begin{lemma}
		\label{lemma CDF-ORB err}
		As $N \rightarrow \infty$, for any non-transmitted message, i.e., $w^{\prime}\neq 1$, and for any $\theta  < 0$, the decoding metric (\ref{decoding metric}) behaves almost surely as
		\begin{align}
			&\lim\limits_{N\to\infty} \frac{1}{N}\ln\mathbb{E}\left\{e^{N \theta \mathsf{D}(w^{\prime})}\Big\vert \underline{\mathsf{T}}\right\}\nonumber\\
			=& \mathbb{E}\left[\ln\left(1+e^{\theta\cdot\bigl\lvert\ln\frac{q^+(\mathsf{Y})}{q^-(\mathsf{Y})}\bigr\rvert}\right)\right] - \ln2.
		\end{align}
	\end{lemma}
	
	With these lemmas, we proceed to prove Theorem \ref{thm:gmi}.
	
	For any $\epsilon > 0$, define the event $\mathcal{U}_{\epsilon}$ as
	\begin{align}
		\mathcal{U}_{\epsilon} &= \Bigg\{\mathsf{D}(1) \geq \frac{1}{2}\int_{\mathcal{R}_1}\Bigl\lvert \ln\frac{q^+(y)}{q^-(y)}\Bigr\rvert q^+(y)\mathrm{d}y\nonumber\\
		&\hspace{5em}+\frac{1}{2}\int_{\mathcal{R}_2}\Bigl\lvert \ln\frac{q^+(y)}{q^-(y)}\Bigr\rvert q^-(y)\mathrm{d}y + \epsilon \Bigg\},\nonumber
	\end{align}
	corresponding to the case where the decoding metric of the transmitted message, $\mathsf{D}(1)$, exceeds its expectation (Lemma \ref{lemma CDF-ORB exp}) by at least $\epsilon$. Consequently, the ensemble average error probability of decoding can be expressed as 
	\begin{align}
		&\text{Pr}[\hat{\mathsf{W}} \neq 1]\nonumber\\
		=\quad&\text{Pr}[\hat{\mathsf{W}} \neq 1\vert \mathcal{U}_\epsilon]\text{Pr}[\mathcal{U}_\epsilon] + \text{Pr}[\hat{\mathsf{W}} \neq 1, \mathcal{U}_\epsilon^c]\nonumber \\
		\leq\quad& \text{Pr}[\mathcal{U}_\epsilon] + \text{Pr}[\hat{\mathsf{W}} \neq 1, \mathcal{U}_\epsilon^c].
		\label{Pr}
	\end{align}\par
	Using Lemmas \ref{lemma CDF-ORB exp} and \ref{lemma CDF-ORB variance}, along with Chebyshev inequality, we obtain that for any $\epsilon > 0$,
	\begin{align}
		\lim\limits_{N\to\infty}&\text{Pr}\Biggl[\mathsf{D}(1) \geq \frac{1}{2}\int_{\mathcal{R}_1}\Bigl\lvert \ln\frac{q^+(y)}{q^-(y)}\Bigr\rvert q^+(y)\mathrm{d}y\nonumber\\
		&\quad +\frac{1}{2}\int_{\mathcal{R}_2}\Bigl\lvert \ln\frac{q^+(y)}{q^-(y)}\Bigr\rvert q^-(y)\mathrm{d}y + \epsilon\Biggr]=0.
	\end{align}
	Therefore, $\text{Pr}[\mathcal{U}_\epsilon]$ can be made arbitrarily small as $N$ tends to infinity.
	
	On the other hand, from the decoding rule (\ref{decoding rule}) together with the union bound, we obtain
	\begin{align}
		& \text{Pr}[\hat{\mathsf{W}} \neq 1,\mathcal{U}_\epsilon^c]\nonumber\\
		\leq \quad &\text{Pr}\Biggl[\exists w^{\prime}\neq 1, \mathsf{D}(w^{\prime})<\frac{1}{2}\int_{\mathcal{R}_1}\Bigl\lvert \ln\frac{q^+(y)}{q^-(y)}\Bigr\rvert q^+(y)\mathrm{d}y\nonumber\\
		&\quad +\frac{1}{2}\int_{\mathcal{R}_2}\Bigl\lvert \ln\frac{q^+(y)}{q^-(y)}\Bigr\rvert q^-(y)\mathrm{d}y + \epsilon\Biggr]\nonumber\\
		\leq \quad & e^{NR} \text{Pr}\Bigg[\mathsf{D}(w^{\prime}) < \frac{1}{2}\int_{\mathcal{R}_1}\Bigl\lvert \ln\frac{q^+(y)}{q^-(y)}\Bigr\rvert q^+(y)\mathrm{d}y\nonumber\\
		&\quad +\frac{1}{2}\int_{\mathcal{R}_2}\Bigl\lvert \ln\frac{q^+(y)}{q^-(y)}\Bigr\rvert q^-(y)\mathrm{d}y + \epsilon\Bigg].
		\label{PrUc}
	\end{align}
	Considering the probability in (\ref{PrUc}) conditioned upon $\underline{\mathsf{T}}$, and applying Chernoff bound, we obtain that for any $N$ and any $\theta< 0$,
	\begin{align}
		&-\frac{1}{N}\ln \text{Pr}\Biggl[\mathsf{D}(w^{\prime}) < \frac{1}{2}\int_{\mathcal{R}_1}\Bigl\lvert \ln\frac{q^+(y)}{q^-(y)}\Bigr\rvert q^+(y)\mathrm{d}y\nonumber\\
		&\quad+\frac{1}{2}\int_{\mathcal{R}_2}\Bigl\lvert \ln\frac{q^+(y)}{q^-(y)}\Bigr\rvert q^-(y)\mathrm{d}y + \epsilon\Bigg\vert \underline{\mathsf{T}} \Biggr]\nonumber\\
		\geq\quad&\theta\cdot\Biggl[\frac{1}{2}\int_{\mathcal{R}_1}\Bigl\lvert \ln\frac{q^+(y)}{q^-(y)}\Bigr\rvert q^+(y)\mathrm{d}y\nonumber\\
		&\quad+\frac{1}{2}\int_{\mathcal{R}_2}\Bigl\lvert \ln\frac{q^+(y)}{q^-(y)}\Bigr\rvert q^-(y)\mathrm{d}y+ \epsilon\Biggr]\nonumber\\
		& -\frac{1}{N}\ln\mathbb{E}\left\{e^{N \theta \mathsf{D}(w^{\prime})}\Big\vert \underline{\mathsf{T}}\right\}.
	\end{align}
	Letting $\epsilon\rightarrow 0, N\rightarrow \infty$, and applying the almost sure limit in Lemma \ref{lemma CDF-ORB err}, we obtain
	\begin{align}
		&\text{Pr}\Biggl[\mathsf{D}(w^{\prime}) < \frac{1}{2}\int_{\mathcal{R}_1}\Bigl\lvert \ln\frac{q^+(y)}{q^-(y)}\Bigr\rvert q^+(y)\mathrm{d}y\nonumber\\
		&\quad +\frac{1}{2}\int_{\mathcal{R}_2}\Bigl\lvert \ln\frac{q^+(y)}{q^-(y)}\Bigr\rvert q^-(y)\mathrm{d}y + \epsilon\Bigg\vert \underline{\mathsf{T}} \Biggr]\nonumber\\
		\leq \quad&\exp\Bigg\{-N\Biggl[\ln2-\mathbb{E}\left[\ln\left(1+e^{\theta\cdot\bigl\lvert\ln\frac{q^+(\mathsf{Y})}{q^-(\mathsf{Y})}\bigr\rvert}\right)\right]\nonumber\\ 
		&\hspace{4em} +\theta\cdot \frac{1}{2}\int_{\mathcal{R}_1}\Bigl\lvert \ln\frac{q^+(y)}{q^-(y)}\Bigr\rvert q^+(y)\mathrm{d}y\nonumber\\
		&\hspace{4em}  +\theta\cdot\frac{1}{2}\int_{\mathcal{R}_2}\Bigl\lvert \ln\frac{q^+(y)}{q^-(y)}\Bigr\rvert q^-(y)\mathrm{d}y\Biggr]\Bigg\}.
		\label{Pr1}
	\end{align}
	
	Then applying the law of total expectation to remove the conditioning in the left side of (\ref{Pr1}) and substituting the right side upper bound into (\ref{PrUc}), we establish that the ensemble average error probability of decoding vanishes asymptotically as $N\rightarrow \infty$, whenever the rate $R$ satisfies
	\begin{align}
		R &< I_{\text{CDF-ORB}} := \ln2 - \inf\limits_{\theta < 0}\Bigg\{\mathbb{E}\left[\ln\left(1+e^{\theta\cdot\bigl\lvert\ln\frac{q^+(\mathsf{Y})}{q^-(\mathsf{Y})}\bigr\rvert}\right)\right]\nonumber\\
		&\quad-\theta\cdot\frac{1}{2}\int_{\mathcal{R}_1}\Bigl\lvert \ln\frac{q^+(y)}{q^-(y)}\Bigr\rvert q^+(y)\mathrm{d}y\nonumber\\
		&\quad-\theta\cdot\frac{1}{2}\int_{\mathcal{R}_2}\Bigl\lvert \ln\frac{q^+(y)}{q^-(y)}\Bigr\rvert q^-(y)\mathrm{d}y\Bigg\}.
		\label{eqn:gmi-alternative}
	\end{align}
	
	Now we prove that $I_{\text{CDF-ORB}}$ is equivalent to the mutual information $I(\mathsf{X};\mathsf{Y})$, i.e., symmetric capacity. By noting that $\bigl\lvert\ln\frac{q^+(y)}{q^-(y)}\bigr\rvert= \ln\frac{q^-(y)}{q^+(y)}$ for $y\in\mathcal{R}_1$ and $\bigl\lvert\ln\frac{q^+(y)}{q^-(y)}\bigr\rvert= \ln\frac{q^+(y)}{q^-(y)}$ for $y\in\mathcal{R}_2$, we can rewrite the mutual information (\ref{channel mutual information}) as
	\begin{align}
		I(\mathsf{X};\mathsf{Y}) &=\ln2-\frac{1}{2}\int_{\mathcal{R}_1}\ln\left(1+e^{\bigl\lvert\ln\frac{q^+(y)}{q^-(y)} \bigr\rvert}\right)q^+(y)\mathrm{d}y\nonumber\\
		&\quad-\frac{1}{2}\int_{\mathcal{R}_2}\ln\left(1+e^{-\bigl\lvert\ln\frac{q^+(y)}{q^-(y)} \bigr\rvert}\right)q^+(y)\mathrm{d}y\nonumber\\
		&\quad-\frac{1}{2}\int_{\mathcal{R}_1}\ln\left(1+e^{-\bigl\lvert\ln\frac{q^+(y)}{q^-(y)} \bigr\rvert}\right)q^-(y)\mathrm{d}y\nonumber\\
		&\quad-\frac{1}{2}\int_{\mathcal{R}_2}\ln\left(1+e^{\bigl\lvert\ln\frac{q^+(y)}{q^-(y)} \bigr\rvert}\right)q^-(y)\mathrm{d}y.
		\label{IXY1}
	\end{align}
	By using the fact that $q(y)=\frac{1}{2}(q^+(y)+q^-(y))$, we can further rewrite (\ref{IXY1}) as
	\begin{align}
		&I(\mathsf{X};\mathsf{Y}) =\ln2-\mathbb{E}\left[\ln\left(1+e^{-\bigl\lvert\ln\frac{q^+(\mathsf{Y})}{q^-(\mathsf{Y})}\bigr\rvert}\right)\right]\nonumber\\
		&\quad-\frac{1}{2}\int_{\mathcal{R}_1}\Bigl\lvert \ln\frac{q^+(y)}{q^-(y)}\Bigr\rvert q^+(y)\mathrm{d}y-\frac{1}{2}\int_{\mathcal{R}_2}\Bigl\lvert \ln\frac{q^+(y)}{q^-(y)}\Bigr\rvert q^-(y)\mathrm{d}y.
		\label{IXY2}
	\end{align}
	
	On the other hand, consider the right side of (\ref{eqn:gmi-alternative}) as a function of $\theta<0$:
	\begin{align}
		&I_{\text{CDF-ORB}} = \sup\limits_{\theta<0}f(\theta),\\
		&f(\theta) := \ln2 - \mathbb{E}\left[\ln\left(1+e^{\theta\cdot\bigl\lvert \ln\frac{q^+(\mathsf{Y})}{q^-(\mathsf{Y})}\bigr\rvert}\right)\right]\nonumber\\
		&+\frac{\theta}{2}\int_{\mathcal{R}_1}\Bigl\lvert \ln\frac{q^+(y)}{q^-(y)}\Bigr\rvert q^+(y)\mathrm{d}y+\frac{\theta}{2}\int_{\mathcal{R}_2}\Bigl\lvert \ln\frac{q^+(y)}{q^-(y)}\Bigr\rvert q^-(y)\mathrm{d}y.
	\end{align}    
	
	Therefore, we obtain the following inequalities:
	\begin{align}
		I_{\text{CDF-ORB}}\geq f(-1)=I(\mathsf{X};\mathsf{Y}).
	\end{align}
	Meanwhile, since $I_{\text{CDF-ORB}}$ is the achievable rate for CDF-ORBGRAND under the symmetric input distribution, while $I(\mathsf{X};\mathsf{Y})$ is the symmetric capacity, Shannon's channel coding theorem dictates that
	\begin{align}
		I_{\text{CDF-ORB}}\leq I(\mathsf{X};\mathsf{Y}).
	\end{align}
	Hence, $I_{\text{CDF-ORB}}=I(\mathsf{X};\mathsf{Y})$, and $f(\theta)$ attains its maximum at $\theta=-1$. This completes the proof.
	\section{Application of CDF-ORBGRAND in BICM}\label{sec:bicm}
	
	High-order modulation has been a prevailing technique in contemporary communication systems, for achieving high spectral efficiency. In order to apply GRAND to high-order modulation, we study BICM systems, where a high-order coded modulation is effectively decomposed into multiple bit channels, with the decoding metric for each bit channel separately computed. Remarkably, we establish that CDF-ORBGRAND is capable of achieving the BICM capacity, which was initially derived in the literature by treating the decomposed bit channels as independent.
	
	\subsection{BICM System Model}\label{subsec:bicm-nonideal}
	
	BICM employs an encoder, a bit interleaver $\pi$, and a binary labeling function $\mu: \left\{+1,-1\right\}^m\rightarrow \mathcal{S}$ that maps blocks of $m$ bits to symbols in a constellation $\mathcal{S}$. Let us start with a codebook of length $mN$ and rate $R$ nats per channel use, containing $\lceil e^{NR}\rceil$ codewords. When message $w$ is selected for transmission, the corresponding codeword is denoted by $\Tilde{\underline{x}}(w)$. The codeword is then passed through the interleaver, and the resulting interleaved codeword is denoted by
	\begin{align}
		\underline{x}(w)=&\pi(\Tilde{\underline{x}}(w))\nonumber\\
		=&[x_{1,1}(w),\cdots,x_{1,m}(w),\cdots,x_{N,1}(w),\cdots,x_{N,m}(w)].\nonumber
	\end{align}
	For notational convenience, in the sequel, $\underline{x}(w)$ will be referred to as the codeword, and the original codeword $\Tilde{\underline{x}}(w)$, which can be recovered from $\underline{x}(w)$ through a de-interleaving operation, will not be needed in the analysis. Since we consider random coding, elements of $\underline{x}(w)$ are i.i.d., uniformly drawn from $\{+1, -1\}$. After modulation, the $m$ bits $x_{i, 1}(w), \ldots, x_{i, m}(w)$ as a block are mapped via the binary labeling function $\mu$ into a symbol $s_i(w)$, and the transmitted symbol sequence is denoted by $$\underline{s}(w)=[s_1(w),\cdots,s_N(w)].$$ 
	The elements of the codeword corresponding to the $j$-th bit of the labeling are denoted by $$\underline{x}_j(w)=[x_{1,j}(w),\cdots,x_{N,j}(w)],$$ for $j = 1, \ldots, m$.	Conversely, we define the inverse mapping of the $j$-th bit of the labeling as $b_j: \mathcal{S}\rightarrow\left\{+1,-1\right\}$; that is, $b_j(s)$ is the codeword bit corresponding to the $j$-th position of the transmitted symbol $s$. The received sequence is denoted by $\underline{\mathsf{Y}}=[\mathsf{Y}_1,\cdots,\mathsf{Y}_N]$. 
	
	We consider a general memoryless channel, where the output pdf of $\mathsf{Y}_i$ under input bit $x_{i,j}=+1$ is denoted by $q_j^+(y)$, and under input bit $x_{i,j}=-1$ is denoted by $q_j^-(y)$, respectively. These pdfs can be evaluated as 
	\begin{gather}
		q_j^+(y)= \frac{1}{\lvert \mathcal{S}_j^+\rvert}\sum\limits_{s \in \mathcal{S}_j^+}p(y\vert s), \quad q_j^-(y)= \frac{1}{\lvert \mathcal{S}_j^-\rvert}\sum\limits_{s \in \mathcal{S}_j^-}p(y\vert s),\nonumber 
	\end{gather}
	where $\mathcal{S}_j^+$ and $\mathcal{S}_j^-$ denote the sets of constellation symbols whose $j$-th bit equals $+1$ and $-1$, respectively. 
	
	Define the LLR $\mathsf{T}_{i,j}=\ln\frac{q_j^+(\mathsf{Y}_i)}{q_j^-(\mathsf{Y}_i)}$ and the reliability vector $$[\lvert \mathsf{T}_{1,1}\rvert,\cdots,\lvert \mathsf{T}_{1,m}\rvert,\cdots,\lvert \mathsf{T}_{N,1}\rvert,\cdots,\lvert \mathsf{T}_{N,m}\rvert].$$ For $i=1,\cdots,N$ and $j=1,\cdots,m$, denote by $\mathsf{R}_{i,j}$ the rank of $\lvert\mathsf{T}_{i,j}\rvert$ among the sorted array consisting of $\left\{\lvert \mathsf{T}_{1,1}\rvert,\cdots,\lvert \mathsf{T}_{1,m}\rvert,\cdots,\lvert \mathsf{T}_{N,1}\rvert,\cdots,\lvert \mathsf{T}_{N,m}\rvert\right\}$, from $1$ (the smallest) to $mN$ (the largest). Let $\Psi_1(t),\cdots,\Psi_m(t)$ be the CDFs of $\lvert \mathsf{T}_{\cdot,1}\rvert,\cdots,\lvert \mathsf{T}_{\cdot,m}\rvert$ for $t\geq 0$, respectively. We further define the average CDF as
	\begin{align}
		\bar{\Psi}(t)=\frac{1}{m}\sum_{j=1}^{m}\Psi_j(t). \label{average CDF}
	\end{align}
	We also introduce the sets $\mathcal{R}_{1,j}=\left\{y\vert q_j^+(y)<q_j^-(y)\right\}$ and $\mathcal{R}_{2,j}=\left\{y\vert q_j^+(y)>q_j^-(y)\right\}$.

	For each $j=1,\cdots,m$, we assume that $q_j^+(y)$ and $q_j^-(y)$ satisfy 
	Assumptions~\ref{assumption:light-tail} and \ref{assumption:poly-bound} 
	introduced in Section~\ref{subsec:system model}. 
	Moreover, the average CDF $\bar{\Psi}(\cdot)$ defined in \eqref{average CDF}, 
	together with its inverse $\bar{\Psi}^{-1}(\cdot)$, is assumed to satisfy 
	Assumption~\ref{assumption:finite-derivative}.

	\subsection{BICM Capacity}
	
\begin{figure}[htbp]
		\centering
		\begin{tikzpicture}[scale=1, transform shape,
			block/.style={rectangle, draw, minimum width=0.6cm, minimum height=0.3cm},
			channel/.style={rectangle, draw, minimum width=1cm, minimum height=0.3cm},
			line/.style={-latex, thick}
			]
			\node[block] (enc) {Encoder};
			\node[channel, text width=3.5cm, right=1.5cm of enc, yshift=1cm] (ch1) {Binary-input channel $1$};
			\node[channel, text width=3.5cm, right=1.5cm of enc, yshift=0.2cm] (ch2) {Binary-input channel $2$};
			\node[channel, text width=3.5cm, right=1.5cm of enc, yshift=-1.5cm] (chm) {Binary-input channel $m$};
			\node[right=0.3cm of enc](lmid1){};
			\draw[thick] (enc.east) -- (lmid1.center);
			\node[right=0.6cm of enc, yshift=0.75cm](lmid2){};
			\draw[line] (lmid1.center) -- (lmid2.center);	
			\node[left=0.05cm of lmid2, yshift=0.05cm] (x) {$x$};
			\node[left=0.5cm of ch1] (lout1) {};
			\draw[thick] (lout1.east) -- (ch1.west);
			\node[left=0.5cm of ch2] (lout2) {};
			\draw[thick] (lout2.east) -- (ch2.west);
			\node[left=0.5cm of chm] (loutm) {};
			\draw[thick] (loutm.east) -- (chm.west);
			\node[right=3.2cm of enc, yshift=-0.5cm] (dots) {$\vdots$};
			\node[right=0.5cm of ch1] (rout1) {};
			\draw[thick] (ch1.east) -- (rout1.west);
			\node[right=0.5cm of ch2] (rout2) {};
			\draw[thick] (ch2.east) -- (rout2.west);
			\node[right=0.5cm of chm] (routm) {};
			\draw[thick] (chm.east) -- (routm.west);
			\node[right=5.8cm of enc, yshift=0.75cm](rmid1){};
			\node[right=6.1cm of enc](rmid2){};
			\draw[thick] (rmid1.center) -- (rmid2.center);
			\node[right=6.7cm of enc](rmid3){};
			\draw[line] (rmid2.center) -- (rmid3.center);
			\node[right=0.05cm of rmid1, yshift=0.05cm] (y) {$y$};
		\end{tikzpicture}
		\caption{Parallel channel model of BICM with ideal interleaving.}
		\label{fig:Parallel channel}
	\end{figure}
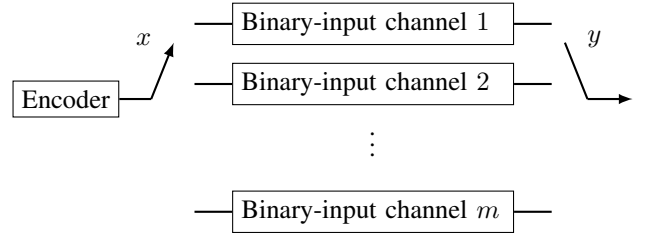

	Analysis of BICM was initially conducted under the assumption of ideal interleaving \cite{caire1998bit}, treating the BICM system model as $m$ mutually independent parallel binary-input memoryless channels, the $j$-th one corresponding to the $j$-th bit of the labeling of the constellation symbols, with output pdf $q^+_j(y)$ under input $x = +1$ and $q^-_j(y)$ under input $x = -1$; see Fig. \ref{fig:Parallel channel} for an illustration.
	
	Based on this parallel channel model, the BICM system achieves the rate, i.e., the BICM capacity, 
	\begin{equation}
		C^{\text{BICM}}=\sum_{j=1}^{m}I(\mathsf{X}_j;\mathsf{Y}),\label{BICM capacity}
	\end{equation}
	where in the $j$-th term of the summation, $\mathsf{X}_j$ is uniformly distributed over $\{+1, -1\}$, and $\mathsf{Y}$ is driven by $q^\pm_j(y)$, as described in the previous paragraph.
	
	Treating the $m$ bits within a symbol as independent is in fact a mismatched decoding rule, which can be expressed as \cite{zehavi19928,i2008bit}
	\begin{align}
		\hat{w} &= \mathop{\arg\max}\limits_{w = 1,\cdots, \lceil e^{NR}\rceil}\prod_{i=1}^{N}\prod_{j=1}^{m}d_j(b_j(s_i(w)),y_i),
		\label{BICM decoder}
	\end{align}
	where the $j$-th bit decoding metric is
	\begin{align}
		d_j(b_j(s)=+1,y)=q_j^+(y),\;d_j(b_j(s)=-1,y)=q_j^-(y).\nonumber
	\end{align}
	Via leveraging the tool of mismatched decoding, it has been shown that the BICM decoding rule (\ref{BICM decoder}) achieves the BICM capacity (\ref{BICM capacity}) even in the absence of ideal interleaving \cite{martinez2009bit}.
	

	\subsection{GRAND and Its Variants for BICM}
	
	After receiving a channel output vector $\underline{y}$, we compute the channel reliability vector $$[\lvert t_{1,1}\rvert,\cdots,\lvert t_{1,m}\rvert,\cdots,\lvert t_{N,1}\rvert,\cdots,\lvert t_{N,m}\rvert]$$ and the corresponding hard-decision vector $$\underline{x}_{\text{hard}}=[\kappa_{1,1}, \cdots,\kappa_{1,m},\cdots, \kappa_{N,1},\cdots,\kappa_{N,m}],$$ where $t_{i,j}=\ln\frac{q_j^+(y_i)}{q_j^-(y_i)}$ is the LLR and $\kappa_{i,j}=\text{sgn}(t_{i,j})$.
	
	GRAND algorithms for BICM have a $2^{mN} \times mN$ ${\pm 1}$-valued matrix $P$ for representing EPs, wherein the $q$-th row is denoted as
	$$P_q=[P_{q,1,1},\cdots,P_{q,1,m},\cdots,P_{q,N,1},\cdots,P_{q,N,m}],$$
	for conducting the $q$-th query: if $P_{q,i,j}=-1$, the sign of $\kappa_{i,j}$ in $\underline{x}_{\text{hard}}$ is flipped, and otherwise it remains unchanged. If $\underline{x}_{\text{hard}}$ after flipping is a codeword, it is declared as the decoder output. If all rows of $P$ are exhausted without finding a codeword, a decoding failure is declared. For ORBGRAND, we arrange the rows of $P$ so that the sum reliability of the $q$-th row, $\sum_{i,j: P_{q,i,j} = -1} r_{i,j}$, is non-decreasing with $q$, where $r_{i,j}$ is the rank of $\lvert t_{i,j}\rvert$.
	
	Similar to Section \ref{subsec:grand}, we have the following unified decoding rule for GRAND algorithms:
	\begin{align}
		\hat{w} = \mathop{\arg\min}\limits_{w = 1,\cdots, \lceil e^{NR}\rceil}\frac{1}{mN}\sum_{i=1}^{N}\sum_{j=1}^{m} \gamma_{i, j}(\underline{y})\cdot \mathbf{1}(\kappa_{i, j} x_{i,j}(w) < 0),
		\label{GRAND under BICM}
	\end{align}
	for which different choices of $\gamma_{i, j}$ correspond to different variants of GRAND under BICM:
	\begin{itemize}
		\item When $\gamma_{i, j}(\underline{y})=1$, (\ref{GRAND under BICM}) is the original GRAND;
		\item When $\gamma_{i, j}(\underline{y})=\lvert t_{i, j} \rvert$,  (\ref{GRAND under BICM}) is SGRAND, and is equivalent to the decoding rule in (\ref{BICM decoder}); see Appendix \ref{BICM decoder equivalence} for a proof of their equivalence;
		\item When $\gamma_{i, j}(\underline{y}) = \frac{r_{i,j}}{mN}$, (\ref{GRAND under BICM}) is ORBGRAND;
		\item When $\gamma_{i, j}(\underline{y}) = \bar{\Psi}(\lvert t_{i, j} \rvert)$, (\ref{GRAND under BICM}) is cdf-GRAND;
		\item When $\gamma_{i, j}(\underline{y})=\bar{\Psi}^{-1}\left(\frac{r_{i,j}}{mN+1}\right)$, where $\bar{\Psi}^{-1}(\cdot)$ denotes the inverse function of $\bar{\Psi}(\cdot)$ given by (\ref{average CDF}), (\ref{GRAND under BICM}) is CDF-ORBGRAND.
	\end{itemize}

	\subsection{Rate Analysis}\label{subsec:rate-bicm}
	
	The following theorem establishes that CDF-ORBGRAND achieves the BICM capacity.
	\begin{theorem}
		\label{Theorem BICM CDF-ORB}
		For the system model in Section \ref{subsec:bicm-nonideal}, CDF-ORBGRAND achieves the BICM capacity, i.e.,
		\begin{align}
			I_{\text{CDF-ORB,BICM}}=C^{\text{BICM}}.
		\end{align}
	\end{theorem}
	\begin{IEEEproof}
		See Appendix \ref{Proof of Theorems BICM}.
	\end{IEEEproof}
	
	For comparison, the following theorem provides the achievable rate of ORBGRAND for BICM.
	\begin{theorem}
		\label{Theorem BICM ORB}
		For the system model in Section \ref{subsec:bicm-nonideal}, ORBGRAND achieves the GMI
		\begin{align}
			I_{\text{ORB,BICM}} &= m\ln2 - \inf\limits_{\theta < 0}\Bigg\{\sum_{j=1}^{m}\Biggl(\int_{0}^{1}\ln(1+e^{\theta t})\mathrm{d}t\nonumber\\
			&-\frac{\theta}{2}\int_{\mathcal{R}_{1,j}}\bar{\Psi}\left(\Bigl\lvert \ln\frac{q_j^+(y)}{q_j^-(y)}\Bigr\rvert\right) q_j^+(y)\mathrm{d}y\nonumber\\
			&-\frac{\theta}{2}\int_{\mathcal{R}_{2,j}}\bar{\Psi}\left(\Bigl\lvert \ln\frac{q_j^+(y)}{q_j^-(y)}\Bigr\rvert\right) q_j^-(y)\mathrm{d}y\Biggr)\Bigg\}\label{ORBGRAND rate BICM}
		\end{align}    
		in nats/channel use.\par
	\end{theorem}
	\begin{IEEEproof}
		See Appendix \ref{Proof of Theorems BICM}.
	\end{IEEEproof}
	

	\subsection{Numerical Results}\label{subsec:numerical-bicm}
	
\begin{figure}[htbp]	
		\centering
		\includegraphics[width=0.49\textwidth]{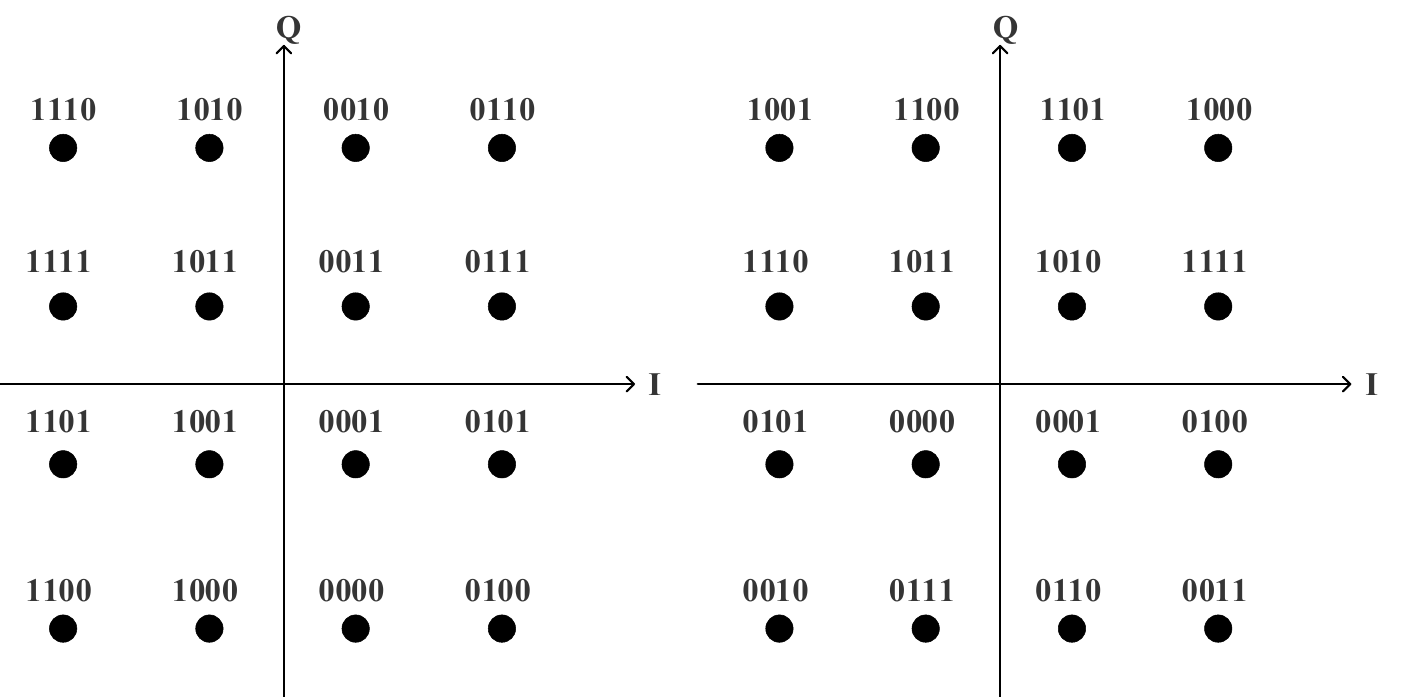}
		\caption{Gray (left) and SP (right) labelings for 16QAM. Here $0$ and $1$ correspond to $-1$ and $+1$, respectively.}
		\label{fig:Gray and SP for 16QAM}
	\end{figure}

	In this subsection, we numerically evaluate $I_{\text{CDF-ORB,BICM}}$ and $I_{\text{ORB,BICM}}$ for QPSK, 8PSK, and 16QAM, over Rayleigh fading channel with receiver-side perfect CSI. For each constellation, we consider both Gray and set-partitioning (SP) labeling schemes. As an illustration, Fig.~\ref{fig:Gray and SP for 16QAM} presents the 16QAM constellation diagrams corresponding to Gray and SP labelings. The channel input–output relationship is given by
	\begin{equation}
		\mathsf{Y} = \mathsf{H}\mathsf{S} + \mathsf{Z},\nonumber
		\label{BICM_model}
	\end{equation}
	where $\mathsf{H} \sim \mathcal{CN}(0, 1)$ is the channel fading coefficient, $\mathsf{S}$ is the channel input corresponding to a constellation symbol, and $\mathsf{Z} \sim \mathcal{CN}(0, 1)$ is the AWGN. The average SNR is thus $\mathbb{E}[\lvert \mathsf{S}\rvert^2]$.
	
\begin{figure*}[t]%
		\centering
		\subfloat[Gray labeling]{
			\includegraphics[width=0.485\linewidth]{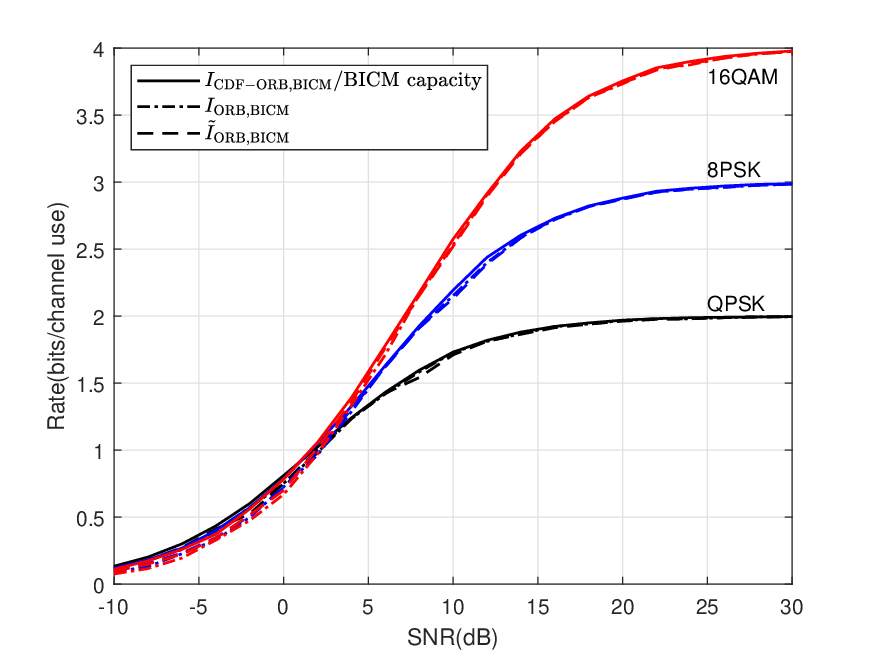}
		}\hfill
		\subfloat[Set-partitioning labeling]{
			\includegraphics[width=0.485\linewidth]{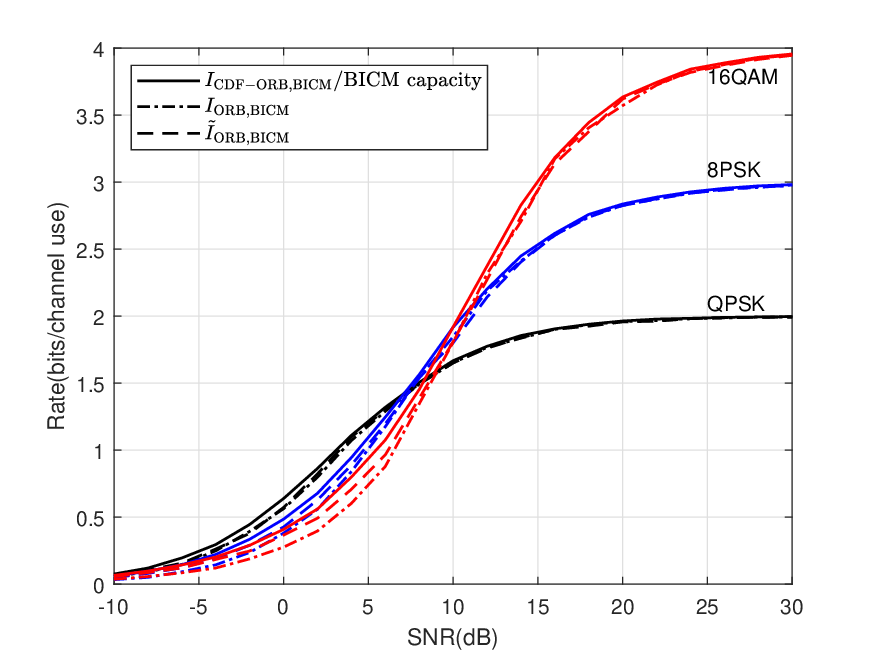}
		}
		\caption{Achievable rates of ORBGRAND, CDF-ORBGRAND, and BICM capacity for QPSK, 8PSK, and 16QAM over Rayleigh fading channel with perfect CSI. Line styles indicate curve type: solid for CDF-ORBGRAND and BICM capacity, dash-dot for ORBGRAND (\ref{ORBGRAND rate BICM}), and dashed for ORBGRAND under the assumption of ideal interleaving (\ref{approx}) in \cite{li2024orbgrand}. Line colors indicate modulation format: black for QPSK, blue for 8PSK, red for 16QAM.}
		\label{fig:Rayleigh}
	\end{figure*}

	The numerical results are shown in Fig. \ref{fig:Rayleigh}. In addition to the achievable rates $I_{\text{CDF-ORB,BICM}}$ and $I_{\text{ORB,BICM}}$ derived in this section, we also plot $\Tilde{I}_{\text{ORB,BICM}}$ from \cite{li2024orbgrand}, which is obtained under the assumption of ideal interleaving; see the end of Appendix~\ref{Proof of Theorems BICM} for details. It can be seen that, under both Gray and SP labelings, ORBGRAND exhibits only a slightly larger gap to the BICM capacity (i.e., $I_{\text{CDF-ORB,BICM}}$) in the low-SNR regime. Under the same conditions, the gap with SP labeling is slightly more pronounced than that with Gray labeling, indicating the impact of the choice of labeling scheme.
	
	Moreover, although the derivation of $\Tilde{I}_{\text{ORB,BICM}}$ relies on the ideal interleaving assumption, we observe that the curves of $\Tilde{I}_{\text{ORB,BICM}}$ are very close to the exact ones of ${I}_{\text{ORB,BICM}}$, indicating that the ideal interleaving assumption is usually sufficient for practical purposes.
	
	\section{Conclusion}\label{Conclusion}
	
	In this work, we first analyze the achievable rates of ORBGRAND and its variant under rank companding, CDF-ORBGRAND, from the perspective of the GMI for general binary-input memoryless channels. A somewhat unexpected yet basic finding is that the achievable rate of CDF-ORBGRAND is identical to the symmetric capacity, indicating that rank companding loses no information asymptotically. Building upon this finding, we extend the analysis to BICM and demonstrate that, CDF-ORBGRAND also achieves the BICM capacity. 
	
	We emphasize that our analytical findings are information-theoretic in nature, which become effective as code lengths grow large. In practical scenarios with finite code lengths, CDF-ORBGRAND still exhibits a performance gap relative to ML decoding. Related to this point, our recent work~\cite{li2026finite} develops a finite-blocklength analysis of ORBGRAND, including an ORBGRAND-specific random-coding union-type achievability bound, a second-order achievable-rate expansion, and the associated normal approximation. Although this analysis is for ORBGRAND rather than CDF-ORBGRAND, it provides a useful basis for extending finite-blocklength analysis to CDF-ORBGRAND and other rank-based GRAND variants. Meanwhile, various practical optimization techniques for ORBGRAND have been proposed to enhance its performance in realistic settings; see, e.g., \cite{condo2021high,duffy2022ordered,liu2022orbgrand,wan2024approaching,wan2025fine}. In consideration of these, evaluating CDF-ORBGRAND under finite code lengths will be the focus of our future work.
	
	\appendices
	
	\section{Proofs of Lemmas \ref{lemma CDF-ORB exp}, \ref{lemma BICM CDF-ORB exp}, and \ref{lemma BICM ORB exp}}\label{Proof of Lemmas 1, 4, 7}
	
	These three lemmas characterize the asymptotic behavior of the expectation of the decoding metric under $w = 1$ as $N \rightarrow \infty$. We only provide the proof of Lemma \ref{lemma CDF-ORB exp}. Lemmas \ref{lemma BICM CDF-ORB exp} and \ref{lemma BICM ORB exp} appear in the proofs of Theorems \ref{Theorem BICM CDF-ORB} and \ref{Theorem BICM ORB}, respectively, in Appendix \ref{Proof of Theorems BICM}. Their proofs are similar to that of Lemma \ref{lemma CDF-ORB exp} and are hence omitted.
	
	For the decoding metric (\ref{decoding metric}) of CDF-ORBGRAND, we have
	\begin{align}
		&\mathbb{E}[\mathsf{D}(1)]=\frac{1}{N}\sum_{i=1}^{N}\nonumber\\
		&\quad\mathbb{E}\left[\Psi^{-1}\left(\frac{\mathsf{R}_i}{N+1}\right)\cdot \mathbf{1}\left(\text{sgn}\left(\mathsf{T}_i\right)\cdot \mathsf{X}_i(1)< 0\right)\right].
		\label{ED1 CDF-ORB}
	\end{align}
	For each summand in (\ref{ED1 CDF-ORB}), we have
	\begin{align}
		&\mathbb{E}\left[\Psi^{-1}\left(\frac{\mathsf{R}_i}{N+1}\right)\cdot \mathbf{1}\left(\text{sgn}\left(\mathsf{T}_i\right)\cdot \mathsf{X}_i(1)< 0\right)\right]\nonumber\\
		=&\frac{1}{2}\mathbb{E}\left[\Psi^{-1}\left(\frac{\mathsf{R}_i}{N+1}\right)\cdot \mathbf{1}\left(\mathsf{T}_i< 0\right)\Bigg\vert \mathsf{X}_i(1)=+1\right]\nonumber\\
		&+\frac{1}{2}\mathbb{E}\left[\Psi^{-1}\left(\frac{\mathsf{R}_i}{N+1}\right)\cdot \mathbf{1}\left(\mathsf{T}_i> 0\right)\Bigg\vert \mathsf{X}_i(1)=-1\right].
		\label{ED1 CDF-ORB two terms}
	\end{align}
	
	Considering the first term in (\ref{ED1 CDF-ORB two terms}) and applying the law of total expectation, we obtain
	\begin{align}
		&\mathbb{E}\left[\Psi^{-1}\left(\frac{\mathsf{R}_i}{N+1}\right)\cdot \mathbf{1}\left(\mathsf{T}_i< 0\right)\Bigg\vert \mathsf{X}_i(1)=+1\right]\nonumber\\
		=&\mathbb{E}\left[\mathbb{E}\left[\Psi^{-1}\left(\frac{\mathsf{R}_i}{N+1}\right)\cdot \mathbf{1}\left(\mathsf{T}_i< 0\right)\Bigg\vert \mathsf{X}_i(1)=+1,\mathsf{Y}_i\right]\right],
	\end{align}
	for which
	\begin{align}
		&\mathbb{E}\left[\Psi^{-1}\Big(\frac{\mathsf{R}_i}{N+1}\Big) \cdot \mathbf{1}(\mathsf{T}_i< 0)\Bigg\vert \mathsf{X}_i(1)=+1,\mathsf{Y}_i=y\right]\nonumber\\
		=&\mathbb{E}\left[\Psi^{-1}\Big(\frac{\mathsf{R}_i}{N+1}\Big) \cdot \mathbf{1}(q^+(y)<q^-(y))\Bigg\vert \mathsf{X}_i(1)=+1,\mathsf{Y}_i=y\right]\nonumber\\
		=& \left\{
		\begin{aligned}
			& \mathbb{E}\left[\Psi^{-1}\Big(\frac{\mathsf{R}_i}{N+1}\Big) \Bigg\vert \mathsf{X}_i(1)=+1,\mathsf{Y}_i=y\right] \, \text{if} \, q^{+}(y)<q^{-}(y),\\
			& 0 \quad\text{else}.\\
		\end{aligned}
		\right.
		\label{total expectation}
	\end{align}
	
	Note that $\mathsf{R}_i$ in (\ref{total expectation}) is the rank of $\lvert t\rvert = \bigl\lvert\ln\frac{q^+(y)}{q^-(y)}\bigr\rvert$ when inserted into a sorted array of $N-1$ i.i.d. samples of $\lvert\mathsf{T}\rvert$, and we denote by $u(y)$ the upper branch of (\ref{total expectation}). Hence we have
	\begin{align}
		&\mathbb{E}\left[\Psi^{-1}\left(\frac{\mathsf{R}_i}{N+1}\right)\cdot \mathbf{1}\left(\mathsf{T}_i< 0\right)\Bigg\vert \mathsf{X}_i(1)=+1\right]\nonumber\\
		=\quad&\int_{\mathcal{R}_1}u(y)q^+(y)\mathrm{d}y.
	\end{align}\par
	
	Similarly, the second term in (\ref{ED1 CDF-ORB two terms}) can be evaluated. So (\ref{ED1 CDF-ORB two terms}) is
	\begin{align}
		&\mathbb{E}\left[\Psi^{-1}\left(\frac{\mathsf{R}_i}{N+1}\right)\cdot \mathbf{1}\left(\text{sgn}\left(\mathsf{T}_i\right)\cdot \mathsf{X}_i(1)< 0\right)\right]\nonumber\\
		=\quad&\frac{1}{2}\int_{\mathcal{R}_1}u(y)q^+(y)\mathrm{d}y+\frac{1}{2}\int_{\mathcal{R}_2}u(y)q^-(y)\mathrm{d}y,
	\end{align}
	which does not depend upon the index $i$, and consequently, (\ref{ED1 CDF-ORB}) becomes
	\begin{align}
		&\mathbb{E}[\mathsf{D}(1)]=\frac{1}{2}\int_{\mathcal{R}_1}u(y)q^+(y)\mathrm{d}y+\frac{1}{2}\int_{\mathcal{R}_2}u(y)q^-(y)\mathrm{d}y.
	\end{align}
	
	The asymptotic behavior of $u(y)$ is given by  Lemma \ref{lemma cdf converge 1_2} in Appendix \ref{Supporting Lemmas}. We thus obtain
	\begin{align} 
		\lim\limits_{N\to\infty} \mathbb{E} [\mathsf{D}(1)] &= \frac{1}{2}\int_{\mathcal{R}_1}\Bigl\lvert \ln\frac{q^+(y)}{q^-(y)}\Bigr\rvert q^+(y)\mathrm{d}y\nonumber\\ &\quad+\frac{1}{2}\int_{\mathcal{R}_2}\Bigl\lvert \ln\frac{q^+(y)}{q^-(y)}\Bigr\rvert q^-(y)\mathrm{d}y.
		\label{limits ED1}
	\end{align} 
	
	\section{Proofs of Lemmas \ref{lemma CDF-ORB variance}, \ref{lemma BICM CDF-ORB variance}, and \ref{lemma BICM ORB variance}}\label{Proof of Lemmas 2, 5, 8}
	
	These three lemmas indicate that the variance of the decoding metric under $w = 1$ asymptotically vanish as $N \rightarrow \infty$. Due to their similarity, we only provide the proof of Lemma \ref{lemma CDF-ORB variance}. Lemmas \ref{lemma BICM CDF-ORB variance} and \ref{lemma BICM ORB variance} appear in the proofs of Theorems \ref{Theorem BICM CDF-ORB} and \ref{Theorem BICM ORB}, respectively, in Appendix \ref{Proof of Theorems BICM}.
	
	Define $\mathsf{W}_i =\Psi^{-1}\left(\frac{\mathsf{R}_i}{N+1} \right)\cdot \mathbf{1}(\text{sgn}(\mathsf{T}_i )\cdot \mathsf{X}_i(1)< 0)$ and $\Tilde{\mathsf{W}}_i = \mathsf{W}_i - \mathbb{E}[\mathsf{W}_i]$. Then for the decoding metric (\ref{decoding metric}) of CDF-ORBGRAND, it follows that
	\begin{align}
		\label{varD1}
		\text{var}\mathsf{D}(1) = \frac{1}{N^2}\sum_{i = 1}^N\sum_{j = 1}^N \mathbb{E}[\Tilde{\mathsf{W}}_i\Tilde{\mathsf{W}}_j].
	\end{align}\par
	
	For $i=j$, we have $\mathbb{E}[\Tilde{\mathsf{W}}_i\Tilde{\mathsf{W}}_j]=\mathbb{E}[\mathsf{W}_i^2]-\mathbb{E}[\mathsf{W}_i]^2$, where $\lim\limits_{N\to\infty} \mathbb{E}[\mathsf{W}_i]$ has been obtained in (\ref{limits ED1}), and following similar steps as those in the proof of Lemma \ref{lemma CDF-ORB exp}, we obtain
	\begin{align}
		\lim\limits_{N\to\infty}\mathbb{E}[\mathsf{W}_i^2] &=\frac{1}{2}\int_{\mathcal{R}_1}{\Big\lvert\ln\frac{q^{+}(y)}{q^{-}(y)}\Big\rvert}^2 q^{+}(y)\mathrm{d}y\nonumber\\
		&\quad+\frac{1}{2}\int_{\mathcal{R}_2}{\Big\lvert\ln\frac{q^{+}(y)}{q^{-}(y)}\Big\rvert}^2 q^{-}(y)\mathrm{d}y.
	\end{align}
	Based on Lemma \ref{lemma finite} in Appendix \ref{Supporting Lemmas}, we establish that  $\lim\limits_{N\to\infty}\mathbb{E}[\Tilde{\mathsf{W}}_i^2]$ remains finite. Since $\text{var}\mathsf{D}(1)$ contains $N$ such terms for $i = 1, \ldots, N$, normalization by $N^2$ in (\ref{varD1}) ensures that their contribution diminishes at a rate of $O\left(\frac{1}{N}\right)$ as $N$ grows without bound.
	
	For $i\neq j$, we have $\mathbb{E}[\tilde{\mathsf{W}}_i\tilde{\mathsf{W}}_j]=\mathbb{E}[\mathsf{W}_i\mathsf{W}_j]-\mathbb{E}[\mathsf{W}_i]\mathbb{E}[\mathsf{W}_j]$. We now turn to the evaluation of $\mathbb{E}[\mathsf{W}_i\mathsf{W}_j]$, which, by definition, can be expressed as (\ref{EWW}), shown at the top of the next page. The first term in (\ref{EWW}) leads to (\ref{total exp2}) and (\ref{total exp3}), shown at the top of the next page. Note that $\mathsf{R}_i$ and $\mathsf{R}_j$ in (\ref{total exp3}) are the ranks resulting from inserting $\lvert t_i\rvert = \bigl\lvert\ln\frac{q^+(y_i)}{q^-(y_i)}\bigr\rvert$ and $\lvert t_j\rvert = \bigl\lvert\ln\frac{q^+(y_j)}{q^-(y_j)}\bigr\rvert$ into a sorted array of $N-2$ i.i.d. samples of $\lvert\mathsf{T}\rvert$, and we denote by $u(y_i,y_j)$ the upper branch of (\ref{total exp3}). Therefore, (\ref{total exp2}) can be rewritten as $\int_{\mathcal{R}_1}\int_{\mathcal{R}_1}u(y_i,y_j)q^+(y_i)\mathrm{d}y_i q^+(y_j)\mathrm{d}y_j$. Applying a similar procedure to the other three terms in (\ref{EWW}), we obtain
	\begin{figure*}[!t] 
		\begin{align}
			\mathbb{E}[\mathsf{W}_i\mathsf{W}_j]&=\mathbb{E}\left[\Psi^{-1}\Big(\frac{\mathsf{R}_i}{N+1}\Big)\Psi^{-1}\Big(\frac{\mathsf{R}_j}{N+1}\Big)\cdot\mathbf{1}(\text{sgn}(\mathsf{T}_i)\cdot \mathsf{X}_i(1)< 0)\mathbf{1}(\text{sgn}(\mathsf{T}_j)\cdot \mathsf{X}_j(1)< 0)\right]\nonumber\\
			&=\frac{1}{4}\mathbb{E}\left[\Psi^{-1}\Big(\frac{\mathsf{R}_i}{N+1}\Big)\Psi^{-1}\Big(\frac{\mathsf{R}_j}{N+1}\Big) \cdot \mathbf{1}(\mathsf{T}_i< 0)\mathbf{1}(\mathsf{T}_j< 0)\Bigg\vert \mathsf{X}_i(1)=+1,\mathsf{X}_j(1)=+1\right]\nonumber\\
			&\quad+\frac{1}{4}\mathbb{E}\left[\Psi^{-1}\Big(\frac{\mathsf{R}_i}{N+1}\Big)\Psi^{-1}\Big(\frac{\mathsf{R}_j}{N+1}\Big) \cdot \mathbf{1}(\mathsf{T}_i> 0)\mathbf{1}(\mathsf{T}_j< 0)\Bigg\vert \mathsf{X}_i(1)=-1,\mathsf{X}_j(1)=+1\right]\nonumber\\
			&\quad+\frac{1}{4}\mathbb{E}\left[\Psi^{-1}\Big(\frac{\mathsf{R}_i}{N+1}\Big)\Psi^{-1}\Big(\frac{\mathsf{R}_j}{N+1}\Big) \cdot \mathbf{1}(\mathsf{T}_i< 0)\mathbf{1}(\mathsf{T}_j> 0)\Bigg\vert \mathsf{X}_i(1)=+1,\mathsf{X}_j(1)=-1\right]\nonumber\\
			&\quad+\frac{1}{4}\mathbb{E}\left[\Psi^{-1}\Big(\frac{\mathsf{R}_i}{N+1}\Big)\Psi^{-1}\Big(\frac{\mathsf{R}_j}{N+1}\Big) \cdot \mathbf{1}(\mathsf{T}_i> 0)\mathbf{1}(\mathsf{T}_j> 0)\Bigg\vert \mathsf{X}_i(1)=-1,\mathsf{X}_j(1)=-1\right],
			\label{EWW}
		\end{align}
		\begin{align}
			&\mathbb{E}\left[\Psi^{-1}\Big(\frac{\mathsf{R}_i}{N+1}\Big)\Psi^{-1}\Big(\frac{\mathsf{R}_j}{N+1}\Big) \cdot \mathbf{1}(\mathsf{T}_i< 0)\mathbf{1}(\mathsf{T}_j< 0)\Bigg\vert \mathsf{X}_i(1)=+1,\mathsf{X}_j(1)=+1\right]\nonumber\\
			=\quad&\mathbb{E}\left[\mathbb{E}\left[\Psi^{-1}\Big(\frac{\mathsf{R}_i}{N+1}\Big)\Psi^{-1}\Big(\frac{\mathsf{R}_j}{N+1}\Big) \cdot \mathbf{1}(\mathsf{T}_i< 0)\mathbf{1}(\mathsf{T}_j< 0)\Bigg\vert \mathsf{X}_i(1)=+1,\mathsf{X}_j(1)=+1,\mathsf{Y}_i,\mathsf{Y}_j\right]\right],
			\label{total exp2}
		\end{align}
	\end{figure*}
	\begin{figure*}
		\begin{align}
			&\mathbb{E}\left[\Psi^{-1}\Big(\frac{\mathsf{R}_i}{N+1}\Big)\Psi^{-1}\Big(\frac{\mathsf{R}_j}{N+1}\Big) \cdot \mathbf{1}(\mathsf{T}_i< 0)\mathbf{1}(\mathsf{T}_j< 0)\Bigg\vert \mathsf{X}_i(1)=+1,\mathsf{X}_j(1)=+1,\mathsf{Y}_i=y_i,\mathsf{Y}_j=y_j\right]\nonumber\\
			=& \left\{
			\begin{aligned}
				&\mathbb{E}\left[\Psi^{-1}\Big(\frac{\mathsf{R}_i}{N+1}\Big)\Psi^{-1}\Big(\frac{\mathsf{R}_j}{N+1}\Big)\Bigg\vert \mathsf{X}_i(1)=+1,\mathsf{X}_j(1)=+1,\mathsf{Y}_i=y_i,\mathsf{Y}_j=y_j\right]\ \text{if}\ q^+(y_i)<q^-(y_i),q^{+}(y_j)<q^-(y_j),\\
				& 0 \quad\text{else}.\\
			\end{aligned}
			\right.
			\label{total exp3}
		\end{align}
	\end{figure*}
	\begin{align}
		\mathbb{E}[\mathsf{W}_i\mathsf{W}_j]&=\frac{1}{4}\int_{\mathcal{R}_1}\int_{\mathcal{R}_1}u(y_i,y_j)q^+(y_i)\mathrm{d}y_i q^+(y_j)\mathrm{d}y_j\nonumber\\
		&\quad+\frac{1}{4}\int_{\mathcal{R}_1}\int_{\mathcal{R}_2}u(y_i,y_j)q^-(y_i)\mathrm{d}y_iq^+(y_j)\mathrm{d}y_j\nonumber\\
		&\quad+\frac{1}{4}\int_{\mathcal{R}_2}\int_{\mathcal{R}_1}u(y_i,y_j)q^+(y_i)\mathrm{d}y_iq^-(y_j)\mathrm{d}y_j\nonumber\\
		&\quad+\frac{1}{4}\int_{\mathcal{R}_2}\int_{\mathcal{R}_2}u(y_i,y_j)q^-(y_i)\mathrm{d}y_iq^-(y_j)\mathrm{d}y_j.
	\end{align}
	
	By applying Lemma \ref{lemma cdf converge 2_2} in Appendix~\ref{Supporting Lemmas}, we obtain that $\lim\limits_{N\to\infty} \mathbb{E}[\Tilde{\mathsf{W}}_i \Tilde{\mathsf{W}}_j] = 0$. Since $\text{var}\mathsf{D}(1)$ contains $N\times(N-1)$ such terms for $i, j = 1, \ldots, N$, normalization by $N^2$ in (\ref{varD1}) ensures that their contribution diminishes as $N$ grows without bound.
	
	Summarizing the cases discussed, we see that $\text{var}\mathsf{D}(1)$ asymptotically vanishes as $N\rightarrow \infty$.
	
	\section{Proofs of Lemmas \ref{lemma CDF-ORB err}, \ref{lemma BICM CDF-ORB err}, and \ref{lemma BICM ORB err}}\label{Proof of Lemmas 3, 6, 9}
	
	\subsection{Proof of Lemma \ref{lemma CDF-ORB err}}
	For any $w^\prime\neq 1$, the conditional expectation of the CDF-ORBGRAND decoding metric (\ref{decoding metric}) satisfies
	\begin{align}
		&\mathbb{E}\left\{e^{N \theta \mathsf{D}(w^{\prime})}\Big\vert \underline{\mathsf{T}}\right\}\nonumber\\
		=\quad&\mathbb{E}\left\{\prod_{i=1}^{N}e^{\theta\Psi^{-1}\left(\frac{\mathsf{R}_i}{N+1}\right)\cdot\mathbf{1}(\text{sgn}(\mathsf{T}_i)\cdot \mathsf{X}_i(w^{\prime})< 0)}\Bigg\vert \underline{\mathsf{T}} \right\}\nonumber\\
		=\quad&\prod_{i=1}^{N}\mathbb{E}\left\{e^{\theta\Psi^{-1}\left(\frac{\mathsf{R}_i}{N+1}\right)\cdot\mathbf{1}(\text{sgn}(\mathsf{T}_i)\cdot \mathsf{X}_i(w^{\prime})< 0)}\Bigg\vert \underline{\mathsf{T}} \right\},
		\label{prod}
	\end{align}
	where we use the fact that $\underline{\mathsf{T}}$ is induced by $\underline{\mathsf{X}}(1)$ and is therefore independent of $\underline{\mathsf{X}}(w^\prime)$. Each term in (\ref{prod}) is
	\begin{align}
		&\mathbb{E}\left\{e^{\theta\Psi^{-1}\left(\frac{\mathsf{R}_i}{N+1}\right)\cdot\mathbf{1}(\text{sgn}(\mathsf{T}_i)\cdot \mathsf{X}_i(w^{\prime})< 0)}\Bigg\vert \underline{\mathsf{T}} \right\}\nonumber \\
		=\quad&\frac{1}{2}\mathbb{E}\left\{e^{\theta\Psi^{-1}\left(\frac{\mathsf{R}_i}{N+1}\right)\cdot\mathbf{1}(\mathsf{T}_i< 0)}\Bigg\vert \underline{\mathsf{T}},\mathsf{X}_i(w^{\prime})=+1 \right\}\nonumber\\
		&+\frac{1}{2}\mathbb{E}\left\{e^{\theta\Psi^{-1}\left(\frac{\mathsf{R}_i}{N+1}\right)\cdot\mathbf{1}(\mathsf{T}_i> 0)}\Bigg\vert \underline{\mathsf{T}},\mathsf{X}_i(w^{\prime})=-1 \right\}\nonumber\\
		=\quad&\frac{1}{2}e^{\theta\Psi^{-1}\left(\frac{\mathsf{R}_i}{N+1}\right)\cdot\mathbf{1}(\mathsf{T}_i< 0)}+\frac{1}{2}e^{\theta\Psi^{-1}\left(\frac{\mathsf{R}_i}{N+1}\right)\cdot\mathbf{1}(\mathsf{T}_i> 0)}\nonumber\\
		=\quad&\frac{1}{2}\left(1+e^{\theta\Psi^{-1}\left(\frac{\mathsf{R}_i}{N+1}\right)}\right),
	\end{align}
	where since $\mathsf{R}_i$ is determined by $\underline{\mathsf{T}}$, the conditional expectation can be suppressed. Therefore, we have
	\begin{align}
		&\frac{1}{N}\ln\mathbb{E}\Bigl\{e^{N \theta \mathsf{D}(w^{\prime})}\Big\vert \underline{\mathsf{T}}\Bigr\}\nonumber\\
		=\quad&\frac{1}{N}\sum_{i=1}^{N}\ln\left(1+e^{\theta\Psi^{-1}\left(\frac{\mathsf{R}_i}{N+1}\right)}\right)-\ln2\nonumber\\
		=\quad&\frac{1}{N}\sum_{n=1}^{N}\ln\left(1+e^{\theta\Psi^{-1}\left(\frac{n}{N+1}\right)}\right)-\ln2,
	\end{align}
	where we use the fact that $\left\{\mathsf{R}_i\right\}_{i=1\cdots,N}$ is a permutation of $\left\{1,\cdots,N\right\}$. Thus, we obtain
	\begin{align}
		&\lim\limits_{N\to\infty}\frac{1}{N}\ln\mathbb{E}\Bigl\{e^{N \theta \mathsf{D}(w^{\prime})}\Big\vert \underline{\mathsf{T}}\Bigr\}\nonumber\\
		=\quad&\lim\limits_{N\to\infty}\frac{1}{N}\sum_{n=1}^{N}\ln\left(1+e^{\theta\Psi^{-1}\left(\frac{n}{N+1}\right)}\right)-\ln2\nonumber\\
		=\quad&\int_{0}^{1}\ln\left(1+e^{\theta\Psi^{-1}(u)}\right)\mathrm{d}u-\ln2\nonumber\\
		=\quad&\int_{0}^{\infty}\ln\left(1+e^{\theta t}\right)\Psi^\prime(t)\mathrm{d}t-\ln2\nonumber\\
		=\quad&\mathbb{E}\left[\ln\left(1+e^{\theta\cdot\bigl\lvert\ln\frac{q^+(\mathsf{Y})}{q^-(\mathsf{Y})}\bigr\rvert}\right)\right] - \ln2,
	\end{align}
	which completes the proof.
	
	\subsection{Proof of Lemma \ref{lemma BICM CDF-ORB err}}
	By following similar steps in the proof of Lemma \ref{lemma CDF-ORB err}, for any $w^\prime \neq 1$, the conditional expectation of the CDF-ORBGRAND decoding metric for BICM in (\ref{BICM decoding metric CDF-ORBGRAND}) satisfies
	\begin{align}
		&\frac{1}{N}\ln\mathbb{E}\Bigl\{e^{N \theta \mathsf{D}(w^{\prime})}\Big\vert \underline{\mathsf{T}}\Bigr\}\nonumber\\
		=\quad&\frac{1}{N}\sum_{i=1}^{N}\sum_{j=1}^{m}\ln\left(1+e^{\frac{\theta}{m}\bar{\Psi}^{-1}\left(\frac{\mathsf{R}_{i,j}}{mN+1}\right)}\right)-m\ln2\nonumber\\
		=\quad&\frac{1}{N}\sum_{n=1}^{mN}\ln\left(1+e^{\frac{\theta}{m}\bar{\Psi}^{-1}\left(\frac{n}{mN+1}\right)}\right)-m\ln2.
	\end{align}
	Thus, we obtain
	\begin{align}
		&\lim\limits_{N\to\infty}\frac{1}{N}\ln\mathbb{E}\Bigl\{e^{N \theta \mathsf{D}(w^{\prime})}\Big\vert \underline{\mathsf{T}}\Bigr\}\nonumber\\
		=\quad&m\int_{0}^{1}\ln\left(1+e^{\frac{\theta}{m}\bar{\Psi}^{-1}(u)}\right)\mathrm{d}u-m\ln2\nonumber\\
		=\quad&m\int_{0}^{\infty}\ln\left(1+e^{\frac{\theta}{m} t}\right)\bar{\Psi}^\prime(t)\mathrm{d}t-m\ln2\nonumber\\
		=\quad&m\int_{0}^{\infty}\ln\left(1+e^{\frac{\theta}{m} t}\right)\cdot \Bigl(\frac{1}{m}\sum_{j=1}^{m}\Psi_j^\prime(t)\Bigr)\mathrm{d}t-m\ln2\nonumber\\
		=\quad&\sum_{j=1}^{m}\int_{0}^{\infty}\ln\left(1+e^{\frac{\theta}{m} t}\right)\Psi_j^\prime(t)\mathrm{d}t-m\ln2\nonumber\\
		=\quad&\sum_{j=1}^{m}\mathbb{E}\Biggl[\ln\Biggl(1+e^{\frac{\theta}{m}\cdot\bigl\lvert\ln\frac{q_j^+(\mathsf{Y})}{q_j^-(\mathsf{Y})}\bigr\rvert}\Biggr)\Biggr] - m\ln2,
	\end{align}
	which completes the proof.
	
	\subsection{Proof of Lemma \ref{lemma BICM ORB err}}
	
	By following similar steps in the proof of Lemma \ref{lemma CDF-ORB err}, for any $w^\prime \neq 1$, the conditional expectation of the ORBGRAND decoding metric for BICM in (\ref{BICM decoding metric ORBGRAND}) satisfies
	\begin{align}
		\frac{1}{N}\ln\mathbb{E}\Bigl\{e^{N \theta \mathsf{D}(w^{\prime})}\Big\vert \underline{\mathsf{T}}\Bigr\}&=\frac{1}{N}\sum_{n=1}^{mN}\ln\left(1+e^{\frac{\theta}{m}\cdot\frac{n}{mN}}\right)-m\ln2\nonumber\\
		&\rightarrow m\int_{0}^{1}\ln(1+e^{\frac{\theta}{m}t})\mathrm{d}t - m\ln2
	\end{align}
	as $N \rightarrow \infty$, which completes the proof.
	
	\section{Equivalence of BICM Decoder (\ref{BICM decoder}) and SGRAND}\label{BICM decoder equivalence}
	
	The decoding rule of SGRAND for BICM can be equivalently expressed as
	\begin{align}
		\hat{w} &= \mathop{\arg\min}\limits_{w = 1,\cdots, \lceil e^{NR}\rceil}\sum_{i=1}^{N}\sum_{j=1}^{m}\nonumber\\ 
		&\quad\Bigl\lvert \ln\frac{q_j^+(y_i)}{q_j^-(y_i)}\Bigr\rvert\cdot \mathbf{1}\left(\text{sgn}\left(\ln\frac{q_j^+(y_i)}{q_j^-(y_i)}\right)\cdot x_{i,j}(w)< 0\right).
		\label{SGRAND equivalent decoder}
	\end{align}
	Likewise, the BICM decoding rule (\ref{BICM decoder}) can be equivalently expressed by
	\begin{align}
		\hat{w} &= \mathop{\arg\max}\limits_{w = 1,\cdots, \lceil e^{NR}\rceil}\sum_{i=1}^{N}\sum_{j=1}^{m}\ln d_j(b_j(s_i(w)),y_i),
		\label{BICM equivalent decoder}
	\end{align}
	where $d_j(+1,y)=q_j^+(y)$ and $d_j(-1,y)=q_j^-(y)$. 
	
	The summation in (\ref{SGRAND equivalent decoder}) can be expanded as
	\begin{align}
		&\sum\limits_{(i,j):x_{i,j}(w)=+1,q_j^+(y_i)<q_j^-(y_i)}[\ln q_j^-(y_i)-\ln q_j^+(y_i)]\nonumber\\
		+&\sum\limits_{(i,j):x_{i,j}(w)=-1,q_j^+(y_i)>q_j^-(y_i)}[\ln q_j^+(y_i)-\ln q_j^-(y_i)],
		\label{sum1}
	\end{align} 
	and the summation in (\ref{BICM equivalent decoder}) is
	\begin{align}
		&\sum\limits_{(i,j):x_{i,j}(w)=+1}\ln q_j^+(y_i)+\sum\limits_{(i,j):x_{i,j}(w)=-1}\ln q_j^-(y_i).
		\label{sum2}
	\end{align}
	
	The sum of (\ref{sum1}) and (\ref{sum2}), after simplification, turns out to be a constant
	\begin{align}
		\sum_{i=1}^{N}\sum_{j=1}^{m}\max\left\{\ln q_j^+(y_i),\ln q_j^-(y_i)\right\}.
	\end{align}
	Therefore, the two decoding rules are equivalent.
	
	\section{Proofs of Theorems \ref{Theorem BICM CDF-ORB} and \ref{Theorem BICM ORB}}\label{Proof of Theorems BICM}
	
	\subsection{Proof of Theorem \ref{Theorem BICM CDF-ORB}}
	
	Based on the unified decoding rule (\ref{GRAND under BICM}), the decoding metric of CDF-ORBGRAND for BICM is given by
	\begin{align}
		\mathsf{D}(w)&=\frac{1}{mN}\sum_{i=1}^{N}\sum_{j=1}^{m}
		\bar{\Psi}^{-1}\left(\frac{\mathsf{R}_{i,j}}{mN+1} \right)
		\cdot\nonumber\\ &\quad\mathbf{1}\left(\text{sgn}\left(\mathsf{T}_{i,j}\right)\cdot \mathsf{X}_{i,j}(w)< 0\right), \quad w=1, \cdots, \lceil e^{NR}\rceil.
		\label{BICM decoding metric CDF-ORBGRAND}
	\end{align}
	
	Under the condition that the transmitted message is $w=1$, the following lemmas characterize some asymptotic properties of the CDF-ORBGRAND decoding metric (\ref{BICM decoding metric CDF-ORBGRAND}). The proofs of these lemmas are provided in the preceding appendices.
	
	\begin{lemma}
		\label{lemma BICM CDF-ORB exp}
		As $N \rightarrow \infty$, the expected value of the decoding metric (\ref{BICM decoding metric CDF-ORBGRAND}) for the transmitted message is given by
		\begin{align}
			\lim\limits_{N\to\infty} \mathbb{E} \mathsf{D}(1) &= \frac{1}{m}\sum_{j=1}^{m}\Biggl(\frac{1}{2}\int_{\mathcal{R}_{1,j}}\Bigl\lvert \ln\frac{q_j^+(y)}{q_j^-(y)}\Bigr\rvert q_j^+(y)\mathrm{d}y\nonumber\\
			&\quad +\frac{1}{2}\int_{\mathcal{R}_{2,j}}\Bigl\lvert \ln\frac{q_j^+(y)}{q_j^-(y)}\Bigr\rvert q_j^-(y)\mathrm{d}y\Biggr).  
		\end{align}     	
	\end{lemma}
	
	\begin{lemma}
		\label{lemma BICM CDF-ORB variance}
		As $N \rightarrow \infty$, the variance of the decoding metric (\ref{BICM decoding metric CDF-ORBGRAND}) for the transmitted message satisfies
		\begin{align}
			\lim\limits_{N\to\infty} \text{var} \mathsf{D}(1) =  0.          
		\end{align}
	\end{lemma}
	
	\begin{lemma}
		\label{lemma BICM CDF-ORB err}
		As $N \rightarrow \infty$, for any non-transmitted message, i.e., $w^{\prime}\neq 1$, and for any $\theta  < 0$, the decoding metric (\ref{BICM decoding metric CDF-ORBGRAND}) behaves almost surely as
		\begin{align}
			& \lim\limits_{N\to\infty} \frac{1}{N}\ln\mathbb{E}\left\{e^{N \theta \mathsf{D}(w^{\prime})}\Big\vert \underline{\mathsf{T}}\right\}\nonumber\\
			&=\sum_{j=1}^{m}\mathbb{E}\Biggl[\ln\Biggl(1+e^{\frac{\theta}{m}\cdot\bigl\lvert\ln\frac{q_j^+(\mathsf{Y})}{q_j^-(\mathsf{Y})}\bigr\rvert}\Biggr)\Biggr] - m\ln2.
		\end{align}
	\end{lemma}
	
	Similar to the proof of Theorem \ref{thm:gmi}, we first define the event $\mathcal{U}_{\epsilon}$ that the decoding metric (\ref{BICM decoding metric CDF-ORBGRAND}) for the transmitted codeword exceeds its asymptotic mean by at least $\epsilon$. This allows us to upper bound the ensemble average error probability of decoding by $\text{Pr}[\mathcal{U}_\epsilon] + \text{Pr}[\hat{\mathsf{W}} \neq 1, \mathcal{U}_\epsilon^c]$. Using Lemmas \ref{lemma BICM CDF-ORB exp} and \ref{lemma BICM CDF-ORB variance} together with Chebyshev inequality, we obtain that $\text{Pr}[\mathcal{U}_\epsilon]\rightarrow 0$ as $N\rightarrow \infty$. On the other hand, by applying Lemma \ref{lemma BICM CDF-ORB err}, we deduce that for any $R$ satisfying 
	\begin{align}
		R&<I_{\text{CDF-ORB,BICM}} :=m\ln2\nonumber\\
		&\quad-\inf\limits_{\theta< 0}\Biggl\{\sum_{j=1}^{m}\Biggl(\mathbb{E}\Bigl[\ln\Bigl(1+e^{\frac{\theta}{m}\cdot\bigl\lvert\ln\frac{q_j^+(\mathsf{Y})}{q_j^-(\mathsf{Y})}\bigr\rvert}\Bigr) \Bigr]\nonumber\\
		&\quad-\frac{\theta}{m}\cdot\frac{1}{2}\int_{\mathcal{R}_{1,j}}\Bigl\lvert\ln\frac{q_j^{+}(y)}{q_j^{-}(y)}\Bigr\rvert q_j^{+}(y)\mathrm{d}y\nonumber\\
		&\quad-\frac{\theta}{m}\cdot\frac{1}{2}\int_{\mathcal{R}_{2,j}}\Bigl\lvert\ln\frac{q_j^{+}(y)}{q_j^{-}(y)}\Bigr\rvert q_j^{-}(y)\mathrm{d}y\Biggr) \Biggr\}
		\label{R}
	\end{align}
	in nats/channel use, the average error probability of decoding asymptotically vanishes as $N \to \infty$.
	
	By setting $\eta=\frac{\theta}{m}$, the right side of (\ref{R}) can be written as
	\begin{align}
		&\quad\sup\limits_{\eta< 0}\Biggl\{\sum_{j=1}^{m}\Biggl(\ln2-\mathbb{E}\Bigl[\ln\Bigl(1+e^{\eta\cdot\bigl\lvert\ln\frac{q_j^+(\mathsf{Y})}{q_j^-(\mathsf{Y})}\bigr\rvert}\Bigr) \Bigr]\nonumber\\
		&+\frac{\eta}{2}\int_{\mathcal{R}_{1,j}}\Bigl\lvert\ln\frac{q_j^{+}(y)}{q_j^{-}(y)}\Bigr\rvert q_j^{+}(y)\mathrm{d}y +\frac{\eta}{2}\int_{\mathcal{R}_{2,j}}\Bigl\lvert\ln\frac{q_j^{+}(y)}{q_j^{-}(y)}\Bigr\rvert q_j^{-}(y)\mathrm{d}y\Biggr) \Biggr\}\nonumber\\
		&= \sup\limits_{\eta< 0}\Bigg\{\sum_{j=1}^{m}f_j(\eta)\Bigg\}.
	\end{align}
	From the proof of Theorem \ref{thm:gmi} in Section \ref{subsec:proof-cdf-rate}, we know that for each $j = 1, \ldots, m$, $f_j(\eta)$ attains its extreme at $\eta=-1$ and $f_j(-1) = I(\mathsf{X}_j;\mathsf{Y})$. Hence, we have the achievable rate of CDF-ORBGRAND for BICM
	\begin{align}
		I_{\text{CDF-ORB,BICM}}=\sup\limits_{\eta< 0}\Big\{\sum_{j=1}^{m}f_j(\eta)\Big\}=\sum_{j=1}^{m}I(\mathsf{X}_j;\mathsf{Y})=C^{\text{BICM}},
	\end{align}
	thereby completing the proof.
	
	\subsection{Proof of Theorem \ref{Theorem BICM ORB}}
	
	Based on the unified decoding rule (\ref{GRAND under BICM}), the decoding metric of ORBGRAND for BICM is given by
	\begin{align}
		\mathsf{D}(w)&=\frac{1}{mN}\sum_{i=1}^{N}\sum_{j=1}^{m} \frac{\mathsf{R}_{i,j}}{mN}\cdot\nonumber\\
		&\quad \mathbf{1}\left(\text{sgn}\left(\mathsf{T}_{i,j}\right)\cdot \mathsf{X}_{i,j}(w)< 0\right), w=1, \cdots, \lceil e^{NR}\rceil.
		\label{BICM decoding metric ORBGRAND}
	\end{align}
	
	Under the condition that the transmitted message is $w=1$, the following lemmas characterize some asymptotic properties of the ORBGRAND decoding metric (\ref{BICM decoding metric ORBGRAND}). The proofs of these lemmas are provided in the preceding appendices.
	
	\begin{lemma}
		\label{lemma BICM ORB exp}
		As $N \rightarrow \infty$, the expected value of the decoding metric (\ref{BICM decoding metric ORBGRAND}) for the transmitted message is given by
		\begin{align}
			\lim\limits_{N\to\infty} \mathbb{E} \mathsf{D}(1) &= \frac{1}{m}\sum_{j=1}^{m}\Bigl(\frac{1}{2}\int_{\mathcal{R}_{1,j}}\bar{\Psi}\Bigl(\Bigl\lvert \ln\frac{q_j^+(y)}{q_j^-(y)}\Bigr\rvert\Bigr) q_j^+(y)\mathrm{d}y\nonumber\\
			&\quad +\frac{1}{2}\int_{\mathcal{R}_{2,j}}\bar{\Psi}\Bigl(\Bigl\lvert \ln\frac{q_j^+(y)}{q_j^-(y)}\Bigr\rvert\Bigr) q_j^-(y)\mathrm{d}y\Bigr).  
		\end{align}     	
	\end{lemma}
	
	\begin{lemma}
		\label{lemma BICM ORB variance}
		As $N \rightarrow \infty$, the variance of the decoding metric (\ref{BICM decoding metric ORBGRAND}) for the transmitted message satisfies
		\begin{align}
			\lim\limits_{N\to\infty} \text{var} \mathsf{D}(1) =  0.          
		\end{align}
	\end{lemma}
	
	\begin{lemma}
		\label{lemma BICM ORB err}
		As $N \rightarrow \infty$, for any non-transmitted message, i.e., $w^{\prime}\neq 1$, and for any $\theta  < 0$, the decoding metric (\ref{BICM decoding metric ORBGRAND}) behaves almost surely as
		\begin{align}
			& \lim\limits_{N\to\infty} \frac{1}{N}\ln\mathbb{E}\left\{e^{N \theta \mathsf{D}(w^{\prime})}\Big\vert \underline{\mathsf{T}}\right\}\nonumber\\		&=m\int_{0}^{1}\ln(1+e^{\frac{\theta}{m}t})\mathrm{d}t - m\ln2.
		\end{align}
	\end{lemma}
	
	Based on these lemmas, we can complete the proof of Theorem \ref{Theorem BICM ORB} following steps similar to those used in the proofs of Theorems \ref{thm:gmi} and \ref{Theorem BICM CDF-ORB}.
	
	As a side remark, we compare ${I}_{\text{ORB,BICM}}$ with the achievable rate derived in our previous work \cite{li2024orbgrand}, where we assumed ideal interleaving (see Fig. \ref{fig:Parallel channel}) and hence the achievable rate is simply the sum of GMIs over the $m$ bit channels, i.e.,
	\begin{align}
		\Tilde{I}_{\text{ORB,BICM}} &= \sum_{j=1}^{m} I_{\text{ORB},j}\nonumber\\
		&=m\ln2-\sum_{j=1}^{m}\inf\limits_{\theta_j < 0}\Bigg\{\int_{0}^{1} \ln(1 + e^{\theta_j t})\mathrm{d}t\nonumber\\
		&\quad-\theta_j\cdot\frac{1}{2}\int_{\mathcal{R}_{1,j}}\Psi_j\Bigl(\Bigl\lvert \ln\frac{q_j^+(y)}{q_j^-(y)}\Bigr\rvert\Bigr) q_j^+(y)\mathrm{d}y\nonumber\\
		&\quad-\theta_j\cdot\frac{1}{2}\int_{\mathcal{R}_{2,j}}\Psi_j\Bigl(\Bigl\lvert \ln\frac{q_j^+(y)}{q_j^-(y)}\Bigr\rvert\Bigr)q_j^-(y)\mathrm{d}y\Bigg\}.
		\label{approx}
	\end{align}
	In contrast to ${I}_{\text{ORB,BICM}}$ given by (\ref{ORBGRAND rate BICM}), here the expression (\ref{approx}) suggests that coding is done separately over different bit channels, rather than across them jointly. When all bit channels have identical laws, i.e., $q_j^{\pm}(y)$ being the same for $j = 1, \ldots, m$, ${I}_{\text{ORB,BICM}}$ and $\Tilde{I}_{\text{ORB,BICM}}$ coincide. Nevertheless, as shown in our numerical results in Section \ref{subsec:numerical-bicm}, their discrepancy is typically negligible.

	\section{Supporting Lemmas}\label{Supporting Lemmas}
	
	\begin{lemma}
		\label{lemma finite}
		If $q^+(y)$ and $q^-(y)$ satisfy Assumptions \ref{assumption:light-tail} and \ref{assumption:poly-bound}, then 
		\begin{align}
			&\frac{1}{2}\int_{\mathcal{R}_1}{\Bigl\lvert\ln\frac{q^{+}(y)}{q^{-}(y)}\Bigr\rvert}^2 q^{+}(y)\mathrm{d}y+\frac{1}{2}\int_{\mathcal{R}_2}{\Bigl\lvert\ln\frac{q^{+}(y)}{q^{-}(y)}\Bigr\rvert}^2 q^{-}(y)\mathrm{d}y\nonumber\\
			&-\left(\frac{1}{2}\int_{\mathcal{R}_1}\Bigl\lvert \ln\frac{q^+(y)}{q^-(y)}\Bigr\rvert q^+(y)\mathrm{d}y+\frac{1}{2}\int_{\mathcal{R}_2}\Bigl\lvert \ln\frac{q^+(y)}{q^-(y)}\Bigr\rvert q^-(y)\mathrm{d}y \right)^2
			\label{integral}
		\end{align}
		is a finite value.
	\end{lemma}
	\begin{IEEEproof}
		Considering the first term in \eqref{integral}, there exists a sufficiently large constant $M$ such that
		\begin{small}
			\begin{align}
				&\int_{\mathcal{R}_1}{\Bigl\lvert\ln\frac{q^{+}(y)}{q^{-}(y)}\Bigr\rvert}^2 q^{+}(y)\mathrm{d}y\label{integral1}\\
				\leq&\int_{\lvert y\rvert< M}{\Bigl\lvert\ln\frac{q^{+}(y)}{q^{-}(y)}\Bigr\rvert}^2 q^{+}(y)\mathrm{d}y+\int_{\lvert y\rvert\geq M}{\Bigl\lvert\ln\frac{q^{+}(y)}{q^{-}(y)}\Bigr\rvert}^2 q^{+}(y)\mathrm{d}y\nonumber\\
				\leq&\int_{\lvert y\rvert< M}{\Bigl\lvert\ln\frac{q^{+}(y)}{q^{-}(y)}\Bigr\rvert}^2 q^{+}(y)\mathrm{d}y+\int_{\lvert y\rvert\geq M}S_2^2(\lvert y\rvert) S_1(\lvert y \rvert)e^{-a\lvert y\rvert}\mathrm{d}y.\label{integral3}
			\end{align}
		\end{small}
		The first term of (\ref{integral3}) is the integral of a bounded function over a bounded interval, so it is finite; the second term of (\ref{integral3}) is also finite, as light-tailed distributions admit all moments. Therefore, (\ref{integral1}) is bounded. Applying analogous argument to the other three terms in (\ref{integral}) shows that all the integrals involved are finite, and hence (\ref{integral}) is finite.
	\end{IEEEproof}
	
	\begin{lemma}
		\label{lemma cdf converge 1_2}
		Suppose that $q_j^+(y)$ and $q_j^-(y)$ satisfy 
		Assumptions~\ref{assumption:light-tail} and \ref{assumption:poly-bound} 
		for all $j=1,\ldots,m$, and that $\bar{\Psi}(\cdot)$, together with its inverse $\bar{\Psi}^{-1}(\cdot)$, satisfies Assumption~\ref{assumption:finite-derivative}. 
		Denote by $\mathsf{R}(v)$ the rank of $v$ when inserted into the sorted array consisting of independent samples 
		$\left\{\lvert \mathsf{T}_{1,1}\rvert,\cdots,\lvert \mathsf{T}_{1,m}\rvert,
		\cdots,\lvert 	\mathsf{T}_{N,1}\rvert,\cdots,\lvert \mathsf{T}_{N,m}\rvert\right\}$, 
		with an arbitrary one of them removed. Then, for every finite $v$ in the support of $\bar{\Psi}(\cdot)$ with $\bar{\Psi}(v)<1$, the following limits hold:
		\begin{align}
			&\lim\limits_{N\to\infty}\frac{\mathbb{E}[\mathsf{R}(v)]}{mN}=\bar{\Psi}(v),\label{rank exp2}\\
			&\lim\limits_{N\to\infty}\frac{\mathbb{E}[{\mathsf{R}(v)}^2]}{m^2N^2}=\bar{\Psi}(v)^2,\label{square rank exp2}\\
			&\lim\limits_{N\to\infty}\mathbb{E}\Big[\bar{\Psi}^{-1}\Big(\frac{\mathsf{R}(v)}{mN+1}\Big)\Big]=v,\label{rank inverse2}\\
			&\lim\limits_{N\to\infty}\mathbb{E}\Big[\Bigl(\bar{\Psi}^{-1}\Big(\frac{\mathsf{R}(v)}{mN+1}\Big)\Bigr)^2\Big]=v^2.\label{square rank inverse2}
		\end{align}
	\end{lemma}
	
	\begin{IEEEproof}
		Without loss of generality, we remove $\lvert \mathsf{T}_{N,m}\rvert$, and the proof holds with very minor modification if we remove any
		other sample. Let
		\[
		p_v=\bar{\Psi}(v),\qquad 
		\mathsf{S}_N=\frac{\mathsf{R}(v)}{mN+1}.
		\]
		
		By the definition of $\mathsf{R}(v)$,
		\begin{align}\label{eqn:rank-v2}
			\mathsf{R}(v)
			=
			1+\sum_{i=1}^{N-1}\sum_{j=1}^{m}
			\mathbf{1}(\lvert\mathsf{T}_{i,j}\rvert <v)
			+\sum_{j=1}^{m-1}\mathbf{1}(\lvert\mathsf{T}_{N,j}\rvert<v).
		\end{align}
		Taking the expectation of \eqref{eqn:rank-v2} gives
		\begin{align}
			\mathbb{E}[\mathsf{R}(v)]
			=
			1+(N-1)\sum_{j=1}^{m}\Psi_j(v)
			+\sum_{j=1}^{m-1}\Psi_j(v).
		\end{align}
		Dividing by $mN$ and letting $N\to\infty$, we obtain
		\begin{align}
			\lim_{N\to\infty}
			\frac{\mathbb{E}[\mathsf{R}(v)]}{mN}
			=
			\frac{1}{m}\sum_{j=1}^{m}\Psi_j(v)
			=
			\bar{\Psi}(v),
		\end{align}
		which proves \eqref{rank exp2}.
		
		Next, since $\mathsf{R}(v)$ is a sum of independent Bernoulli random variables plus one, 
		\[
		\operatorname{var}[\mathsf{R}(v)]=O(N).
		\]
		Therefore,
		\[
		\operatorname{var}[\mathsf{S}_N]
		=
		\frac{\operatorname{var}[\mathsf{R}(v)]}{(mN+1)^2}
		=
		O\left(\frac{1}{N}\right).
		\]
		Moreover, from the preceding expression for $\mathbb{E}[\mathsf{R}(v)]$, we have $\mathbb{E}[\mathsf{S}_N]\to p_v$. Hence,
		\begin{align}
			\mathbb{E}\left[(\mathsf{S}_N-p_v)^2\right]
			=
			\operatorname{var}[\mathsf{S}_N]
			+
			\left(\mathbb{E}[\mathsf{S}_N]-p_v\right)^2
			\to0.
		\end{align}
		Thus, $\mathsf{S}_N\to p_v$ in mean square and hence also in probability. In particular, $\mathbb{E}[\mathsf{S}_N^2]\to p_v^2$. Since
		\[
		\mathsf{S}_N^2
		=
		\frac{\mathsf{R}(v)^2}{(mN+1)^2},
		\]
		and $(mN+1)^2/(m^2N^2)\to1$, it follows that
		\begin{align}
			\lim_{N\to\infty}
			\frac{\mathbb{E}[\mathsf{R}(v)^2]}{m^2N^2}
			=
			p_v^2
			=
			\bar{\Psi}(v)^2,
		\end{align}
		which proves \eqref{square rank exp2}.
		
		It remains to prove \eqref{rank inverse2} and \eqref{square rank inverse2}. Let $G(u)=\bar{\Psi}^{-1}(u)$, $0\leq u<1$. By Assumption~\ref{assumption:finite-derivative},
		\[
		p_v<1,\qquad G(p_v)=v,
		\]
		and $G(\cdot)$ is continuous at $p_v$. Since $\mathsf{S}_N\to p_v$ in probability, the continuous mapping theorem gives
		\[
		G(\mathsf{S}_N)\to v
		\]
		in probability, and consequently $G(\mathsf{S}_N)^2\to v^2$ in probability.
		
		We next justify the required uniform integrability. By Assumptions~\ref{assumption:light-tail} and \ref{assumption:poly-bound}, the average reliability distribution has a sufficiently light tail, so that its inverse satisfies
		\begin{align}\label{eqn:A}
			G\left(1-\frac{1}{n}\right)
			\leq
			A(1+(\log n)^d)
		\end{align}
		for some constants $A>0$, $d\geq1$, and all sufficiently large $n$. Since $\mathsf{R}(v)\leq mN$, we have
		\[
		\mathsf{S}_N
		\leq
		\frac{mN}{mN+1}
		=
		1-\frac{1}{mN+1}.
		\]
		Using the monotonicity of $G(\cdot)$, it follows that
		\begin{align}\label{eqn:bound}
			G(\mathsf{S}_N)
			\leq
			G\left(1-\frac{1}{mN+1}\right)
			\leq
			A(1+(\log N)^d),
		\end{align}
		for some constant $A>0$, possibly larger than that in \eqref{eqn:A}.
		
		Furthermore, since $p_v<1$, choose $\delta>0$ such that $p_v+\delta<1$. On the event $\{\mathsf{S}_N\leq p_v+\delta\}$,
		\[
		G(\mathsf{S}_N)^2
		\leq
		G(p_v+\delta)^2
		<\infty.
		\]
		On the complementary event, Hoeffding's inequality gives
		\[
		\Pr(\mathsf{S}_N>p_v+\delta)\leq e^{-c_\delta N}
		\]
		for some $c_\delta>0$. Combining this with the preceding logarithmic-polynomial bound \eqref{eqn:bound} on $G(\mathsf{S}_N)$, we obtain
		\begin{align}
			&\mathbb{E}\left[
			G(\mathsf{S}_N)^2
			\mathbf{1}\{\mathsf{S}_N>p_v+\delta\}
			\right]\nonumber\\
			&\leq
			A^2(1+(\log N)^d)^2 e^{-c_\delta N}
			\to 0.
		\end{align}
		Therefore, $\{G(\mathsf{S}_N)^2\}_{N\geq1}$ is uniformly integrable. Consequently, $\{G(\mathsf{S}_N)\}_{N\geq1}$ is also uniformly integrable.
		
		Since $G(\mathsf{S}_N)\to v$ in probability and $\{G(\mathsf{S}_N)\}_{N\geq1}$ is uniformly integrable, the standard uniform-integrability convergence theorem implies $\mathbb{E}[G(\mathsf{S}_N)]\to v$. Similarly, since $G(\mathsf{S}_N)^2\to v^2$ in probability and $\{G(\mathsf{S}_N)^2\}_{N\geq1}$ is uniformly integrable, we obtain $\mathbb{E}[G(\mathsf{S}_N)^2]\to v^2$. These are exactly \eqref{rank inverse2} and \eqref{square rank inverse2}. The proof is complete.
	\end{IEEEproof}
	
	\begin{lemma}
		\label{lemma cdf converge 2_2}
		Suppose that $q_j^+(y)$ and $q_j^-(y)$ satisfy 
		Assumptions~\ref{assumption:light-tail} and \ref{assumption:poly-bound} 
		for all $j=1,\ldots,m$, and that $\bar{\Psi}(\cdot)$, together with its inverse 
		$\bar{\Psi}^{-1}(\cdot)$, satisfies Assumption~\ref{assumption:finite-derivative}. 
		Denote by $\mathsf{R}(v_a)$ and $\mathsf{R}(v_b)$ the ranks of $v_a$ and $v_b$, respectively, 
		when inserted into the sorted array consisting of independent samples 
		$\left\{\lvert \mathsf{T}_{1,1}\rvert,\cdots,\lvert \mathsf{T}_{1,m}\rvert,
		\cdots,\lvert 	\mathsf{T}_{N,1}\rvert,\cdots,\lvert \mathsf{T}_{N,m}\rvert\right\}$, 
		with arbitrary two of them removed. Then, for every finite $v_a$ and $v_b$ in the support of $\bar{\Psi}(\cdot)$ with 
		$\bar{\Psi}(v_a)<1$ and $\bar{\Psi}(v_b)<1$, the following limits hold:
		\begin{align}
			&\lim\limits_{N\to\infty}
			\frac{\mathbb{E}[\mathsf{R}(v_a)\mathsf{R}(v_b)]}{m^2N^2}
			=\bar{\Psi}(v_a)\bar{\Psi}(v_b),\label{two ranks2}\\
			&\lim\limits_{N\to\infty}\mathbb{E}\Bigg[
			\bar{\Psi}^{-1}\Big(\frac{\mathsf{R}(v_a)}{mN+1}\Big)
			\bar{\Psi}^{-1}\Big(\frac{\mathsf{R}(v_b)}{mN+1}\Big)
			\Bigg]
			=v_av_b.\label{two ranks inverse2}
		\end{align}
	\end{lemma}
	
	\begin{IEEEproof}
		Without loss of generality, we remove $\lvert\mathsf{T}_{N,m-1}\rvert$ and 
		$\lvert\mathsf{T}_{N,m}\rvert$, and the proof holds with very minor modification if we remove any other two samples. Let
		\[
		p_a=\bar{\Psi}(v_a),\qquad p_b=\bar{\Psi}(v_b),
		\]
		and define
		\[
		\mathsf{S}_{N,a}=\frac{\mathsf{R}(v_a)}{mN+1},\qquad
		\mathsf{S}_{N,b}=\frac{\mathsf{R}(v_b)}{mN+1}.
		\]
		Following the same argument as that in Lemma~\ref{lemma cdf converge 1_2}, we have
		\[
		\mathsf{S}_{N,a}\to p_a,\qquad
		\mathsf{S}_{N,b}\to p_b
		\]
		in mean square. Hence, by the Cauchy--Schwarz inequality,
		\begin{align}
			&\left|\mathbb{E}[\mathsf{S}_{N,a}\mathsf{S}_{N,b}]-p_ap_b\right|\nonumber\\
			&\leq
			\sqrt{\mathbb{E}\left[(\mathsf{S}_{N,a}-p_a)^2\right]}
			\sqrt{\mathbb{E}\left[(\mathsf{S}_{N,b}-p_b)^2\right]}\nonumber\\
			&\quad
			+p_a\left|\mathbb{E}[\mathsf{S}_{N,b}-p_b]\right|
			+p_b\left|\mathbb{E}[\mathsf{S}_{N,a}-p_a]\right|
			\to0.
		\end{align}
		Hence,
		\begin{align}
			\mathbb{E}[\mathsf{S}_{N,a}\mathsf{S}_{N,b}]\to p_ap_b.
		\end{align}
		Since
		\begin{align}
			\mathsf{S}_{N,a}\mathsf{S}_{N,b}
			=
			\frac{\mathsf{R}(v_a)\mathsf{R}(v_b)}{(mN+1)^2}
		\end{align}
		and $(mN+1)^2/(m^2N^2)\to1$, we obtain
		\begin{align}
			\lim_{N\to\infty}
			\frac{\mathbb{E}[\mathsf{R}(v_a)\mathsf{R}(v_b)]}{m^2N^2}
			=
			p_ap_b
			=
			\bar{\Psi}(v_a)\bar{\Psi}(v_b),
		\end{align}
		which proves \eqref{two ranks2}.
		
		Let $G(u)=\bar{\Psi}^{-1}(u)$. By Assumption~\ref{assumption:finite-derivative} and the continuous mapping theorem,
		\[
		G(\mathsf{S}_{N,a})G(\mathsf{S}_{N,b})\to v_av_b
		\]
		in probability. Moreover, by the same uniform-integrability argument as that in Lemma~\ref{lemma cdf converge 1_2}, 
		$\{G(\mathsf{S}_{N,a})^2\}_{N\geq1}$ and 
		$\{G(\mathsf{S}_{N,b})^2\}_{N\geq1}$ are uniformly integrable. Since
		\[
		\left|G(\mathsf{S}_{N,a})G(\mathsf{S}_{N,b})\right|
		\leq
		\frac{G(\mathsf{S}_{N,a})^2+G(\mathsf{S}_{N,b})^2}{2},\]
		$\{G(\mathsf{S}_{N,a})G(\mathsf{S}_{N,b})\}_{N\geq1}$ is also uniformly integrable. Therefore,
		\[
		\mathbb{E}\left[
		G(\mathsf{S}_{N,a})G(\mathsf{S}_{N,b})
		\right]\to v_av_b,
		\]
		which proves \eqref{two ranks inverse2}. The proof is complete.
	\end{IEEEproof}

	\bibliographystyle{IEEEtran}
	\bibliography{ref.bib}
\end{document}